\documentclass[a4paper,11pt]{article}
\pdfoutput=1
\usepackage{dcolumn}
\usepackage{bm}
\usepackage{graphicx}
\usepackage{amssymb,amsmath}
\usepackage{multirow}
\usepackage{booktabs}
\usepackage{cite,color,url}
\usepackage[dvipsnames]{xcolor}
\def\linkcolor{cyan!70!black}

\usepackage[
colorlinks=true
,urlcolor=\linkcolor
,anchorcolor=\linkcolor
,citecolor=\linkcolor
,filecolor=\linkcolor
,linkcolor=\linkcolor
,menucolor=\linkcolor
,linktocpage=true
,pdfproducer=medialab
,pdfa=true
]{hyperref}

\usepackage{slashed}
\usepackage{epsfig,psfrag,rotating,soul}
\usepackage{rotfloat}
\usepackage[font={small}]{caption}
\usepackage{colortbl}

\evensidemargin \oddsidemargin
\allowdisplaybreaks

\let\OLDthebibliography\thebibliography
\renewcommand\thebibliography[1]{
  \OLDthebibliography{#1}
  \setlength{\parskip}{0pt}
  \setlength{\itemsep}{0pt plus 0.3ex}
}

\definecolor{mygray}{gray}{0.85} 
\definecolor{myblue}{cmyk}{0.65, 0.37, 0.0, 0.19}

\begin{document}

\begin{titlepage}

\thispagestyle{empty}

\def\thefootnote{\fnsymbol{footnote}}

\begin{flushright}
IFT-UAM/CSIC-23-107
\end{flushright}

\vspace*{1cm}

\begin{center}

\begin{Large}
\textbf{\textsc{Probing new physics with charge asymmetries in \\[.25em] 2~same-sign leptons plus jets final states at the LHC}}
\end{Large}

\vspace{1cm}

{\sc
Ernesto~Arganda$^{1, 2}$%
\footnote{{\tt \href{mailto:ernesto.arganda@uam.es}{ernesto.arganda@uam.es}}}%
,  Leandro~Da~Rold$^{3}$%
\footnote{{\tt \href{mailto:daroldl@ib.edu.ar}{daroldl@ib.edu.ar}}}%
, Aurelio~Juste$^{4,5}$%
\footnote{{\tt \href{mailto:juste@ifae.es}{juste@ifae.es}}}%
, Anibal~D.~Medina$^{2}$%
\footnote{{\tt \href{mailto:anibal.medina@fisica.unlp.edu.ar}{anibal.medina@fisica.unlp.edu.ar}}} \\%
and Rosa~M.~Sand\'a Seoane$^{1}$%
\footnote{{\tt \href{mailto:r.sanda@csic.es}{r.sanda@csic.es}}}%

}

\vspace*{.7cm}

{\sl
$^1$Departamento de Física Teórica and Instituto de F\'{\i}sica Te\'orica UAM-CSIC, \\
Universidad Autónoma de Madrid, Cantoblanco, 28049 Madrid, Spain

\vspace*{0.1cm}

$^2$IFLP, CONICET - Dpto. de F\'{\i}sica, Universidad Nacional de La Plata, \\ 
C.C. 67, 1900 La Plata, Argentina

\vspace*{0.1cm}

$^3$Centro At\'omico Bariloche, Instituto Balseiro and CONICET, \\
Av. Bustillo 9500, 8400, S. C. de Bariloche, Argentina

\vspace*{0.1cm}

$^4$Institut de F\'{\i}sica d’Altes Energies (IFAE), Edifici Cn, Facultat de Ciencies, \\
Universitat Aut\`onoma de Barcelona, E-08193 Bellaterra, Barcelona, Spain

\vspace*{0.1cm}

$^5$Institució Catalana de Recerca i Estudis Avançats (ICREA), E-08010 Barcelona, Spain
}

\end{center}

\vspace{0.1cm}

\renewcommand*{\thefootnote}{\arabic{footnote}}
\setcounter{footnote}{0}

\begin{abstract}
\noindent
We study the impact of new physics models in the charge asymmetry defined for LHC final states consisting of two same-sign leptons (2SS$l$, with $l= e, \mu$) plus jets ($N_\text{jets}\geq2$), with a center-of-mass energy of $\sqrt{s}=13$ TeV, where the main SM contribution is $t\bar{t}W$ production. Concretely, we consider three different new physics sources for the charge asymmetries: a heavy neutral scalar/pseudoscalar arising from the general two Higgs doublet model, an effective theory with dimension-6 four-quark operators, and a simplified $R$-parity-violating supersymmetric model with electroweakino production (higgsino-like or wino-like). We propose measuring the charge asymmetries differentially with respect to several kinematic observables, and inclusively/exclusively with the number of $b$-tagged jets in the final state ($N_b\geq\{1, 2, 3\}$). Results are compared with the SM prediction using the $\chi^2$ criteria, expressing sensitivity in terms of the normal Gaussian significance. We show that some of the proposed new physics scenarios can be probed at the LHC even for the already recorded integrated luminosity of 139 fb$^{-1}$. Finally, we also estimate the prospects for the potential LHC sensitivity to the considered new physics models in its high-luminosity phase.
\end{abstract}


\def\thefootnote{\arabic{footnote}}
\setcounter{page}{0}
\setcounter{footnote}{0}

\end{titlepage}

\tableofcontents


\section{Introduction}
\label{intro}

After a decade of operation of the Large Hadron Collider (LHC), the search for new phenomena that could signal the breakdown of the Standard Model (SM) continues at full steam. The large recorded data sets of proton--proton ($pp$) collisions at the energy frontier by the ATLAS and CMS experiments, enhanced through continuous improvements in object reconstruction and calibration, event simulation, and analysis techniques, and supported by state-of-the-art theoretical predictions, allow a vibrant program of direct searches for new particles, as well as a comprehensive program of precision measurements of SM processes. 

One of the most promising final-state signatures at the LHC involves the production of several light leptons (electrons or muons) in association with jets. On the one hand, many beyond-the-SM (BSM) scenarios predict such signatures and, on the other hand, the presence of multiple leptons allow efficient triggering on the events and becomes a powerful handle to suppress multijet backgrounds. Of particular interest is the two same-sign leptons (2SS$l$, with $l= e, \mu$) plus jets final state, which in many LHC studies offers the best compromise between signal yield and background level. After the requirement of at least one $b$-tagged jet, the SM background is dominated by $t\bar{t}W$ production, with additional contributions from $t\bar{t}Z/\gamma^*$, $t\bar{t}H$, $tqZ/\gamma^*$, and diboson production. Depending on the lepton selection criteria applied, backgrounds with non-prompt leptons or jets misidentified as leptons, can also be substantial. The availability of large-statistics data sets, such as those collected during Run 2 of the LHC by the ATLAS and CMS Collaborations, amounting to about 140 fb$^{-1}$ of $pp$ collisions at $\sqrt{s}=13$~TeV, makes the 2SS$l$ channel the primary target of several high-profile measurements~\cite{ATLAS:2019nvo,CMS:2019rvj,CMS:2020mpn,ATLAS:2020hpj,ATLAS:2021kqb,CMS:2023zdh,ATLAS:2023ajo,CMS:2023ftu} and searches (see e.g. Refs.~\cite{ATLAS:2022rws, ATLAS:2023tlp,CMS:2020cpy, ATLAS:2023lfr,CMS:2023xyc}). 

Among the challenges these physics analyses had to face, there was the apparent mismodelling of the $t\bar{t}W$ background, whose cross section was found to be underestimated by factors $\sim$1.3--1.7 with respect to the state-of-art theoretical cross section at the time (NLO+NLL)~\cite{LHCHiggsCrossSectionWorkingGroup:2016ypw}, and the concern whether the $t\bar{t}W$ kinematics was adequately modelled by the Monte Carlo simulation. In response to this situation, a number of recent developments are helping consolidate the theoretical understanding of the $t\bar{t}W$ background, culminating with the calculation of its cross section at NNLO~\cite{Maltoni:2015ena,Dror:2015nkp,Frederix:2017wme,Broggio:2019ewu,Kulesza:2020nfh,Frederix:2020jzp,Bevilacqua:2020pzy,Denner:2020hgg,Bevilacqua:2020srb,FebresCordero:2021kcc,Frederix:2021agh,Bevilacqua:2021tzp,Buonocore:2023ljm}. On the experimental side, the ATLAS and CMS Collaborations have recently performed precise measurements of the total $t\bar{t}W$ cross section~\cite{ATLAS:2023gon,CMS:2022tkv}, which still come out a bit high compared to the NNLO prediction, confirming a trend that was already observed in previous measurements with a smaller data set~\cite{Aaboud:2019njj,Sirunyan:2017uzs}. In addition, for the first time the ATLAS Collaboration has performed a detailed set of $t\bar{t}W$ differential cross-section measurements~\cite{ATLAS:2023gon}, which can be used to validate the kinematic modelling by Monte Carlo (MC) generators, although the measurements are still limited by the data statistics. 

One of the unique observables that can be exploited in the 2SS$l$ final state in $pp$ collisions is the charge asymmetry, $A^{+/-}$, defined as the difference between the numbers of events with two positively charged leptons and with two negatively-charged leptons, divided by their sum. Most backgrounds contributing to the 2SS$l$ final state are charge symmetric and thus have zero charge asymmetry. In contrast, $t\bar{t}W$ production has a strong charge asymmetry, with about twice more events with two positively charged leptons than with two negatively charged leptons. Interestingly, this charge asymmetry is very well known theoretically, since it mainly depends on the $u$- and $d$-quark parton distribution functions (PDF), and thus is quite robust against higher-order QCD effects~\cite{Frixione:2015zaa}. In addition, it can also be measured experimentally with small systematic uncertainties, which largely cancel in the ratio. This raises the interesting possibility to exploit differential measurements of the charge asymmetry in $t\bar{t}W$ production as a robust observable to probe for BSM effects. In fact, the first such measurements have been recently performed by the ATLAS Collaboration~\cite{ATLAS:2023gon}. In this paper we explore this idea and evaluate the potential of such measurements to probe some illustrative new physics signals that could contaminate the $t\bar{t}W$ measurement.

This paper is organized as follows: in Section~\ref{ttW} we present the charge asymmetry, define the observables we are interested in and briefly describe the tools we use for our simulations; in Section~\ref{BSM} we study the charge asymmetry in several extensions of the SM: in the presence of new scalars, in an effective theory with four-quark operators and in an $R$-parity-violating supersymmetric model; in Section~\ref{sec-hllhc} we study the prospects for the high luminosity LHC in those extensions of the SM and we conclude in Section~\ref{sec-conclusions}.

\section{Charge asymmetry in \texorpdfstring{$t \bar t W$}{ttW}  production}
\label{ttW}

As previously discussed, $t \bar t W$ is the main SM contribution to the 2SS$l$+jets final state and has charge-asymmetric production in $pp$ collisions. As such, it constitutes the main SM background in a charge asymmetry measurement. 

The charge asymmetry at the parton level is defined as 
\begin{equation}
A^{+/-}_\textrm{parton}=\frac{\sigma^+-\sigma^-}{\sigma^++\sigma^-}, 
\label{asymmetry_parton}
\end{equation}
where $\sigma^+$ ($\sigma^-$) denotes the $t\bar{t}W^+$ ($t\bar{t}W^-$) cross section. To illustrate the small theoretical uncertainty on the parton-level inclusive charge asymmetry, we reproduce here the prediction obtained by the ATLAS Collaboration at $\sqrt{s}=13$~TeV in Ref.~\cite{ATLAS:2023gon}, using a higher-order calculation from the {\sc Sherpa} MC generator: $A^{+/-}=0.322 \pm 0.003\,(\textrm{scale}) \pm 0.007\,(\textrm{PDF})$. At the reconstructed level, the charge asymmetry can be computed similarly, both inclusively and differentially, using the predicted yields for events with two positively charged leptons ($N^{++}$) and two negatively charged leptons ($N^{--}$):

\begin{equation}
A^{+/-}_\textrm{reco}=\frac{N^{++}-N^{--}}{N^{++}+N^{--}}.
\label{asymmetry}
\end{equation}
\noindent In the remainder of this paper we will be considering the differential charge asymmetry at the reconstructed level, denoted $A^{+/-}$ for simplicity, which will be computed with the aid of MC simulations.

The simulated sample for $t \bar t W$ production was generated with {\sc MadGraph5\_aMC@NLO} \cite{Alwall:2014hca} at NLO in QCD with the NNPDF2.3 NLO PDF set \cite{Ball:2012cx}, and at $\sqrt{s}=13$~TeV . As in Ref. \cite{ATLAS:2019nvo}, we applied two scaling factors to the inclusive cross-section prediction to include missing QCD and electroweak corrections: a 1.1 factor to account for NLO QCD corrections to $t \bar t W+1$-jet \cite{Alwall:2014hca}, and a 1.09 factor to account for sub-leading NLO electroweak corrections \cite{Frederix:2017wme}. We obtained a total yield of 710 fb, in agreement with the prediction used in Ref.~\cite{ATLAS:2019nvo}. Spin correlations in $t \bar t W$ production have sizable effects already at LO in QCD~\cite{Maltoni:2014zpa}, and were included in our sample simulation through {\sc Madspin}~\cite{Artoisenet:2012st, Frixione:2007zp}, for decaying the $t/\bar{t}$ quark and the $W^{+/-}$ boson while preserving the spin orientation. Events were processed with {\sc Pythia} \cite{Sjostrand:2014zea,Sjostrand:2007gs} for parton showering and hadronization, and {\sc Delphes} \cite{deFavereau:2013fsa} for fast detector simulation, using the default ATLAS card. We used the Monash tune \cite{Skands:2014pea} of {\sc Pythia} to reproduce the expected $t \bar t W$ $N_\text{jets}$ distribution in Ref.~\cite{ATLAS:2019nvo} as faithfully as possible.

Following the selection cuts for the 2SS$l$ final state in association with jets in Ref.~\cite{ATLAS:2019nvo}, events are required to have two same-charge very tight light leptons with $p_T>20$ and $|\eta|<2/2.5$ for $e/\mu$, and no hadronically decaying $\tau$-lepton candidates ($\tau_\text{had}$; they must have $p_T>25$ and $|\eta|<2.5$). We require an invariant mass of the dilepton system, $m_\text{inv}(l,l)$, of at least 12 GeV. Jets are required to satisfy $p_T>25$~GeV and $|\eta|<2.5$, and we also require $N_\text{jets}\geq2$. Jets are $b$-tagged using $p_T$-dependent parameterizations depending on jet flavor ($b$, $c$, or light) that correspond to a working point with an average efficiency of 70\% for $b$-quark jets in $t\bar{t}$ events.
We set as default the inclusive $1 b$-jet selection ($N_{b}\geq1$) but we also probe the exclusive $1$ and 2 $b$-jet selection ($N_{b}={1,2}$), as well as the inclusive 3 $b$-jet selection ($N_{b}\geq3$). Event selection criteria is summarized in Table~\ref{table:cuts}.

\begin{table}[]
\centering
\begin{tabular}{@{}c@{}}
\toprule
\multicolumn{1}{c}{Selection cuts}             \\ 
\hline
2SSl ($p_T>20$ GeV $\&$ $|\eta|<2/2.5$ for $e/\mu$)            \\ 
No additional $e/\mu$ candidates ($p_T>10$ GeV $\&$ $|\eta|<2.47/2.5$ for $e/\mu$) \\
No $\tau_\text{had}$ candidates ($p_T>25$ GeV $\&$ $|\eta|<2.5$)           \\  
$N_\text{jets}\geq2$ ($p_T>25$ GeV $\&$ $|\eta|<2.5$) \\ 
$m_\text{inv}(l,l)>12$ GeV                                 \\ 
$N_{b}\geq1$ \\
\bottomrule
\end{tabular}
\caption{Event selection criteria for the 2SSl plus jets final state. Jets containing $b$-hadrons must be first classified as jets.}
\label{table:cuts}
\end{table}

Along with this work, we consider differential charge asymmetries in terms of different kinematic observables, such us $N_\text{jets}$, $N_{b}$, $H_{T_\text{jets}}$ (defined as the scalar sum of the $p_T$ of all jets), $H_{T_\text{lep}}$ (defined as the scalar sum of the $p_T$ of the two leptons), $m_\text{inv}(l,l)$, $\Delta\eta(l,l)$, $\Delta\Phi(l,l)$, and Eff$_\text{mass}$ (defined as the scalar sum of the $p_T$ of all the objects, including the missing transverse energy). In the next section, we report MC uncertainties for the SM theoretical prediction ($t \bar t W$ production only), while we assume statistical uncertainties are dominant in the expected yield for the SM plus new physics signal scenarios. 

We also simulated other irreducible SM backgrounds ($t\Bar{t}H$, $t\Bar{t}(Z/\gamma^{*})$ and $W(Z/\gamma^{*})$) in the 2SS$l$ final state. It is assumed that these background processes are subtracted out in the $t\Bar{t}W$ charge asymmetry measurement, and thus they do not affect the central measured value, but they do contribute to the estimated statistical uncertainty. For all contributions, we used {\sc MadGraph5\_aMC@NLO} for matrix level generation: $t\Bar{t}H$ was generated  at NLO in QCD with the NNPDF2.3 NLO PDF set, while $t\Bar{t}(Z/\gamma^{*})$ and $W(Z/\gamma^{*})$ were generated LO with the NNPDF2.3 LO PDF set \cite{Ball:2012cx}. For the last two processes we considered the decay $Z/\gamma^{*} \to l^+l^-$. For $W(Z/\gamma^{*})$ we also included contributions from $W(Z/\gamma^{*})+j$ and $W(Z/\gamma^{*})+jj$, using the default matching scheme included in {\sc MadGraph5\_aMC@NLO}. Events were also processed through the {\sc Pythia} and {\sc DELPHES} interfaces. In each case, we applied normalization factors to obtain the same ratios with the total $t \bar t W$ yield as the ones reported in Ref.~\cite{ATLAS:2019nvo}. 



\section{Charge asymmetries from new physics}
\label{BSM}

In order to show the strength of the charge asymmetry observable in distinguishing new physics from SM background, in this section we consider several extensions of the SM and study the production of multilepton plus jets final states, focusing on the two same sign leptons (2SS$l$, with $l=e,\mu$) in association with jets ($N_\text{jets}\geq 2$) signal. 

We study three different new physics sources for the charge asymmetries: heavy neutral scalars or pseudoscalars within the framework of the general two Higgs doublet model (g2HDM); an effective field theory with four-quark operators of dimension six; and a simplified $R$-parity-violating supersymmetric model with higgsino-like or wino-like neutralino/chargino production~\footnote{There are numerous new physics sources that can produce charge asymmetries, such as the type-II seesaw model with scalar triplets or the Zee model considered in Ref.~\cite{Babu:2022ycv}.}. Some of these frameworks have been considered in the ATLAS analysis of Ref.~\cite{ATLAS:2023tlp}. The models should therefore be considered as illustrative examples, but the importance and novelty of this work lies in the differential study of the charge asymmetry and we will show how powerful it can be in probing new physics in some of these scenarios. As will be seen later for example, in some of these scenarios the differential charge asymmetry peaks at large values of the jet multiplicity, while in others it peaks at low jet multiplicity. These type of features  can be exploited to distinguish new physics from SM contributions and may also help in distinguishing different new physics scenarios from each other.

Since the only valence quarks of the proton are up and down quarks, the new physics interactions with them contribute to the numerator of the asymmetry, whereas interactions with the sea quarks, being charge symmetric, contribute to the denominator and dilute $A^{+/-}$. Moreover, the contributions of sea quarks are strongly suppressed by their proton PDFs, compared with the light quark ones.\footnote{As an example, we have checked that for g2HDM and the effective theory, for equal couplings with up and charm quarks, the cross section of processes contributing to $A^{+/-}$  initiated by the former are $\sim{\cal O}(300)$ larger than those initiated by the later.}  
Thus in what follows we  assume that new physics couples only to up-quark and quarks of the third generation. Another motivation for avoiding interactions with quarks of second generation is that turning on simultaneous interactions with quarks of first and second generations generically induces flavor transitions that have very stringent bounds~\cite{Isidori:2010kg,ParticleDataGroup:2022pth}.

The ATLAS Collaboration has reported a best fit value of the $t\bar t W$ cross section of $890\pm 80$ fb~\cite{ATLAS:2023gon}, whereas the SM prediction from Ref.~\cite{Frederix:2021agh} is $722^{+70}_{-78}$ (scale) $\pm 7$ (PDF) fb. Thus, in the new physics models below we consider benchmark points exceeding the SM cross section by 20$\%$ and 50$\%$, that roughly correspond to the ratio between the central values, and the ratio considering the deviations of 1$\sigma$, respectively. For our analysis assuming 139 fb$^{-1}$, we consider the inclusive 1 $b$-jet selection only, since other $b$-jet selections are less sensitive to new physics effects. Only in Section~\ref{sec-hllhc} at high luminosity we consider other $b$-jet selections.

In the following, for the figures presenting the differential charge asymmetry, the SM $t\Bar{t}W$ prediction will include only uncertainties from the limited MC statistics available to calculate it (shown as a blue band). On the other hand, the charge asymmetry corresponding to the sum of SM $t\Bar{t}W$ plus new physics, will include the expected statistical uncertainty for its measurement at the assumed integrated luminosity (shown as a red band). In this study, the effect of systematic uncertainties in the charge asymmetry measurement is neglected, given their sub-leading importance in the recent ATLAS measurement~\cite{ATLAS:2023gon}.

\subsection{The general two Higgs doublet model}
\label{g2HDM}

\begin{figure}[t!]
\begin{center}
\begin{tabular}{cc}
\includegraphics[width=0.35\textwidth]{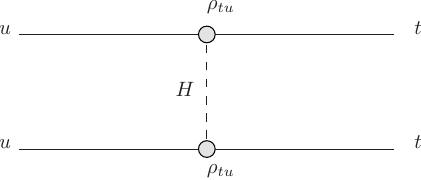}\hspace{1cm} &
\includegraphics[width=0.35\textwidth]{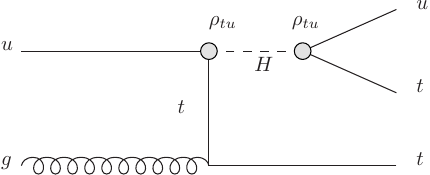} \\
(a)\hspace{1cm} & (b) \\
 & \\
\includegraphics[width=0.35\textwidth]{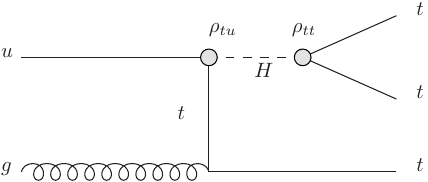}\hspace{1cm} &
\includegraphics[width=0.35\textwidth]{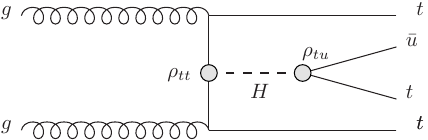} \\
(c)\hspace{1cm} & (d) \\
 & \\
\end{tabular}
\includegraphics[width=0.35\textwidth]{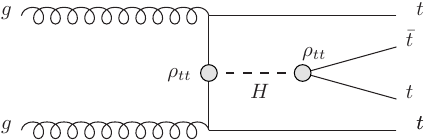}\\
(e)
\caption{Representative signal diagrams in the g2HDM for the exotic CP-even Higgs $H$ that contribute to the 2SS$l$ plus jets final state, where $q=u$ is the up-type quark. Diagrams (a), (b), and (c) and their conjugates give large contributions to the charge asymmetries.}\label{g2HDMdiags}
\end{center}
\end{figure}

Two Higgs doublet models are one of the simplest extensions of the SM~\cite{Lee:1973iz,Branco:2011iw}. Besides simplicity, they can offer solutions to some long standing problems, as the baryon asymmetry of the universe, since they can offer enough new sources of CP violation~\cite{Trodden:1998qg}, they are required in supersymmetric extensions of the SM~\cite{Haber:1984rc}, being indirectly related with the solution of the hierarchy problem and dark matter, as well as in Peccei-Quin solutions of the CP problem of QCD~\cite{Peccei:1977hh,Kim:1986ax} and in some composite Higgs models with non-minimal cosets~\cite{Mrazek:2011iu}. Although these theories usually contain other fields, in many cases the lightest state is a second Higgs doublet such that at low energies the other states can be integrated-out.

We study in this subsection  the general two Higgs doublet model and concentrate on the interactions of the neutral Higgs sector with the SM fermions which are described by the following Lagrangian~\cite{Davidson:2005cw, Altunkaynak:2015twa, Hou:2020ciy, Hou:2020chc}:
\begin{equation}
\mathcal{L}_f=-\frac{1}{\sqrt{2}}\sum_{f=u,d,l}\bar{f}\left[(-\lambda^{f}_{ij} \sin\gamma + \rho^{f}_{ij} \cos\gamma)h+(\lambda^{f}_{ij} \cos\gamma + \rho^{f}_{ij} \sin\gamma)H-i\ {\rm sgn}(Q_f) \rho^{f}_{ij} A\right]P_R f_j + h.c. ,
\end{equation}  
where $h$ and $H$ are the light and heavy CP-even Higgs bosons respectively, $A$ is the CP-odd Higgs boson,  $P_R=(1+\gamma_5)/2$ is the right handed chirality projector, $i,j=1,2,3$ are generation indices that are summed over, $Q_f$ is the electric charge of fermion $f$ and $\gamma$ is the mixing angle in the Higgs CP even sector. The matrices $\lambda^{f}_{ij}\equiv \sqrt{2} m^f_i \delta_{ij}/v$ are the usual Yukawa couplings, whereas $ \rho^{f}_{ij}$  are in general non-diagonal matrices in flavor space which are taken to be real but not necessarily Hermitian, such that CP is conserved in the Higgs sector. Henceforth, we work in the alignment limit $ \cos\gamma\to 0$~\footnote{The alignment limit can obtained for example when one of the quartics involved in the mixing of $h$ and $H$ vanishes~\cite{Davidson:2005cw}. } in which the heavy exotic Higgs bosons ($H$ and $A$) couple with strength $ \rho^{f}_{ij}$ and $ i\rho^{f}_{ij}\gamma_5$  to Dirac fermions respectively, whereas the light CP-even Higgs boson $h$ behaves as the SM-like Higgs boson and couples accordingly. Since only the light Higgs boson carries all of the EW vacuum expectation value (vev), $H$ does not couple linearly to massive SM gauge bosons. Two Higgs doublet models are known to generate dangerous tree-level flavor changing neutral currents (FCNCs), requiring symmetries for their cancellation. Without specification of the ultimate reason suppressing FCNCs, in the present work we will assume that the second Higgs interacts with up type quarks only, avoiding tree-level contributions to mixing of down type mesons. Furthermore, we only consider the exotic couplings involving the top and up quarks to be non-vanishing, $\rho_{tt}, \rho_{tu}, \rho_{ut}$, avoiding leading order contributions to $D$-meson mixing. It is also worth emphasizing that in the alignment limit low energy FCN Higgs (FCNH) effects tend to be suppressed without the necessity to invoke natural flavor conservation, which would imply vanishing FCNH couplings~\cite{Hou:2020ciy}. Thus additional Yukawa couplings that mix different flavors are quite general and their phenomenological implications should be studied. Following Ref.~\cite{ATLAS:2023tlp} we study the production and decay modes considered in that ATLAS analysis as shown in Fig.~\ref{g2HDMdiags}. 

We show the results of our simulations for an LHC center-of-mass energy of $\sqrt{s}=13$ TeV with a total integrated luminosity of $\mathcal{L}=139$ fb$^{-1}$, for the case in which $A$ is decoupled from the theory~\footnote{Due to the appearance of two exotic couplings in the signal diagrams, the CP nature of the scalar becomes irrelevant when considering them one at a time.} and thus only $H$ contributes to the signal. We used the 2HDM model from Ref.~\cite{Degrande:2014vpa} located in the repository from {\sc FeynRules}  and via the interface {\sc MadGraph5\_aMC@NLO}+{\sc Pythia}+{\sc Delphes} we obtained the number of events in the positively and negatively charge branches, respectively. Spin correlations were taken into account via the {\sc MadGraph5\_aMC@NLO} decay chain syntax. In order to reduce the dimensionality of the parameter space, we choose for simplicity $\rho_{tu} = \rho_{ut}$. The strongest experimental constraint comes from the total number of events produced for the final states considered and from bounds three-top-quark production~\cite{ATLAS:2023ajo} and, therefore, for each mass analyzed we only consider values of the couplings such that the cross section to the final states are approximately 50$\%$ the SM contribution provided by $t\bar{t}W$, unless otherwise stated. In this way the total number of events provided by the new physics are buried under the uncertainties of the SM $t\Bar{t}W$ measurement. Given that neither $A$ nor $H$ couple to massive SM gauge bosons, the only other single production mode is gluon fusion which, due to previous considerations, tends to be suppressed for light masses for which $\rho_{tt}<1$. This same effect and the absence of couplings to charge gauge bosons imply that the decay width into diphotons is also small in comparison with a Higgs boson of the same mass and SM couplings. We verified that, indeed, we are several orders of magnitude below the latest bounds provided by spin-0 searches decaying into diphoton final states~\cite{ATLAS:2021uiz}. The presence of $H$ and $A$, and the flavor structure of the couplings considered, also provide a novel production mode for two exotic Higgs bosons ($HH$, $AA$ or even $HA$) from up-type quark initial states exchanging a top-quark in the t-channel via the $\rho_{tu}$ coupling. The subsequent decay of the exotic Higgs bosons into top pairs and/or light jets provides a charge symmetric contribution which could potentially impact on the asymmetry generated from Fig.~\ref{g2HDMdiags}. We calculated the cross section for this process $\sigma_{HH}\sim \mathcal{O}(1)$ fb, providing a much smaller contribution than the ones from Fig.~\ref{g2HDMdiags} that are of order $\sim 200$ fb, implying that we can safely neglect it.

\begin{figure}[t!]
\begin{center}
\includegraphics[width=0.45\textwidth]{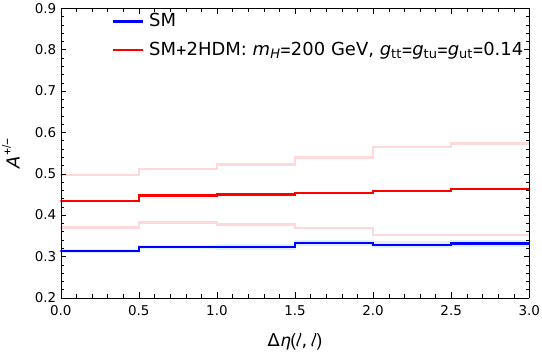}
\includegraphics[width=0.45\textwidth]{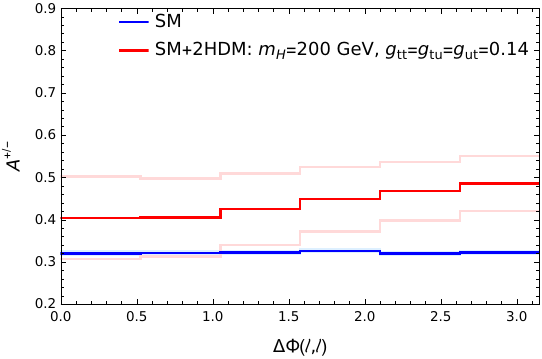}
\vspace{1mm}
\includegraphics[width=0.45\textwidth]{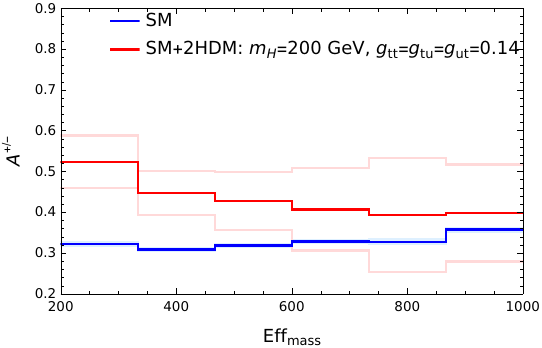}
\includegraphics[width=0.45\textwidth]{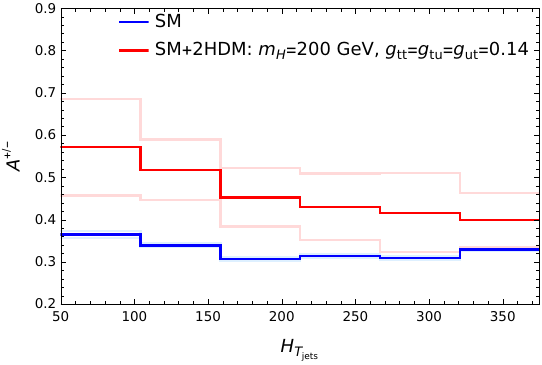}
\vspace{1mm}
\includegraphics[width=0.45\textwidth]{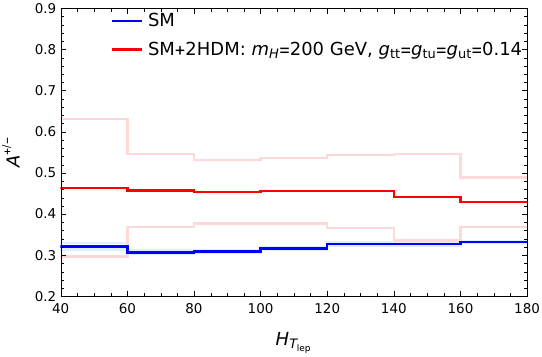}
\includegraphics[width=0.45\textwidth]{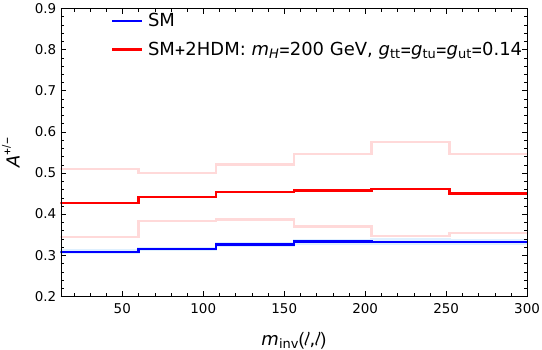}
\vspace{1mm}
\includegraphics[width=0.45\textwidth]{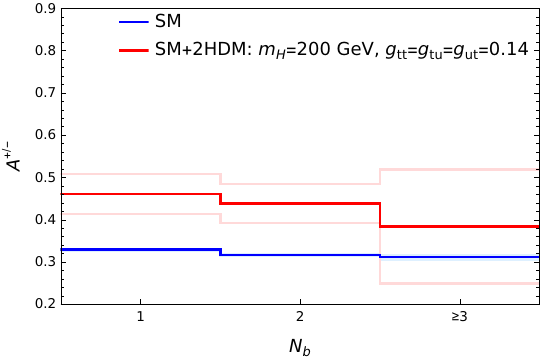}
\includegraphics[width=0.45\textwidth]{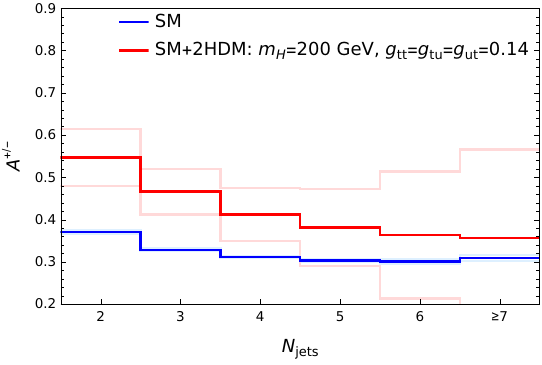}
    \caption{Charge asymmetry distributions for $m_H = 200$ GeV and $\rho_{tt}=\rho_{tu}=0.14$.}
    \label{g2HDM200A}
\end{center}
\end{figure}

\begin{figure}[t]
\begin{center}
-\includegraphics[width=0.45\textwidth]{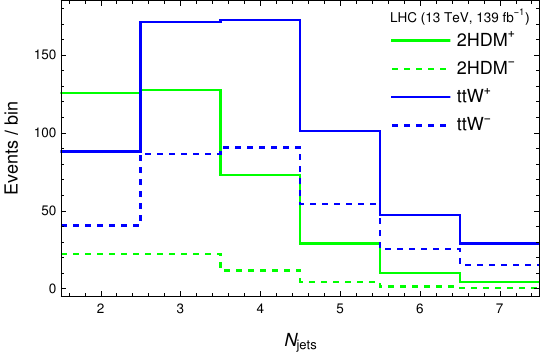}
    \caption{Number of events as function of $N_\text{jets}$ for $m_H = 200$ GeV and $\rho_{tt}=\rho_{tu}=0.14$. }
    \label{g2HDM200events}
\end{center}
\end{figure}

\begin{figure}[t!]
\begin{center}
\includegraphics[width=0.45\textwidth]{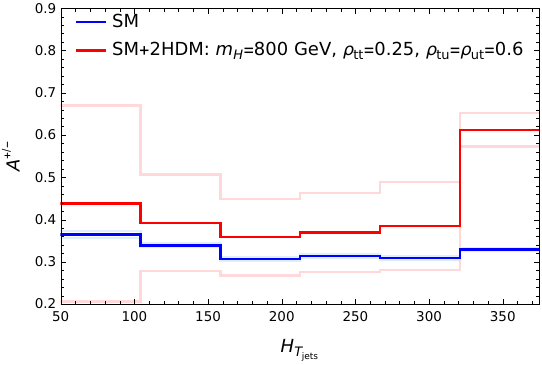}
\includegraphics[width=0.45\textwidth]{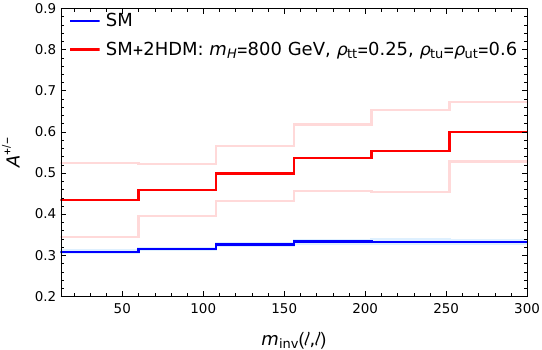}
\vspace{1mm}
\includegraphics[width=0.45\textwidth]{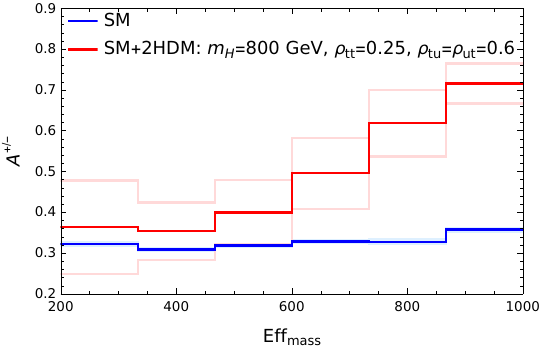}
\includegraphics[width=0.45\textwidth]{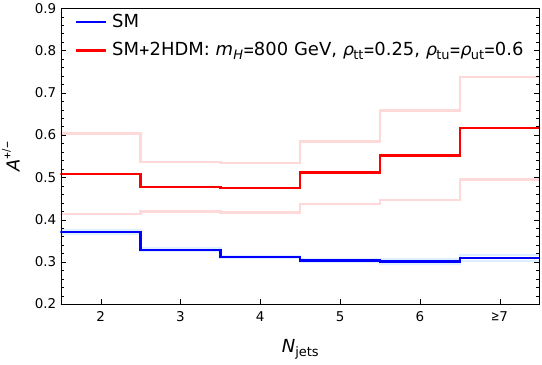}
\vspace{1mm}
\includegraphics[width=0.45\textwidth]{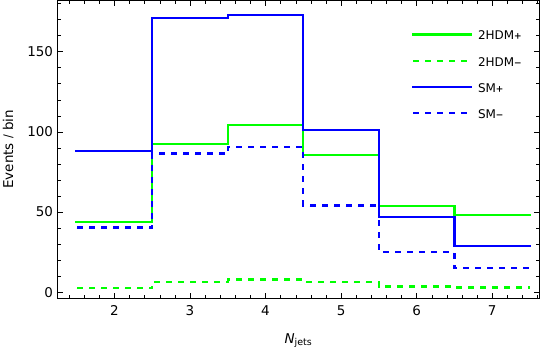}
    \caption{Asymmetry distributions and number of events as function of $N_\text{jets}$ with a scalar $m_H=800$~GeV, $\rho_{tt}=0.25$ and $\rho_{tu}=0.6$.}
    \label{g2HDM400A}
\end{center}
\end{figure}

\begin{figure}[t!]
\begin{center}

\includegraphics[width=0.45\textwidth]{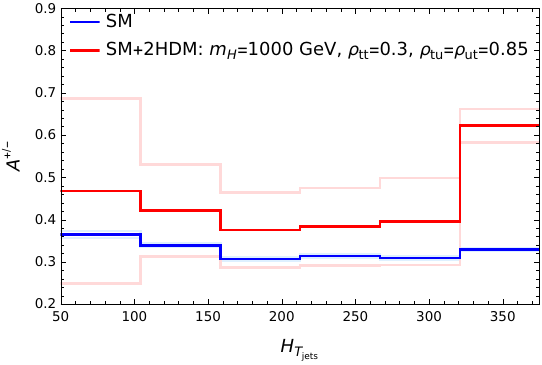}
\includegraphics[width=0.45\textwidth]{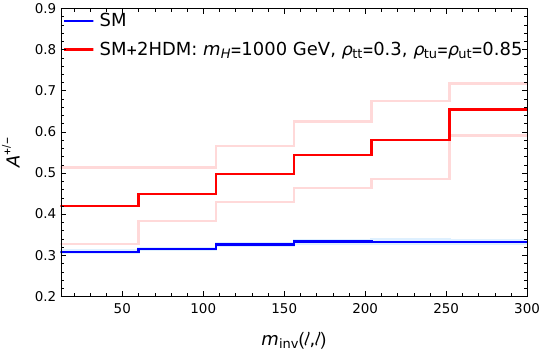}
\vspace{1mm}
\includegraphics[width=0.45\textwidth]{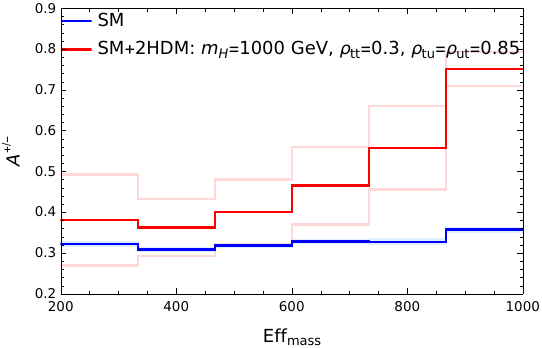}
\includegraphics[width=0.45\textwidth]{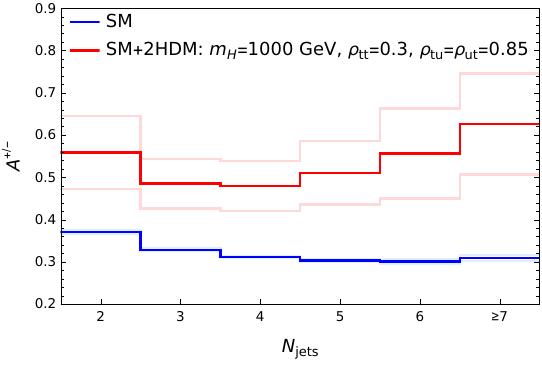}
\vspace{1mm}
\includegraphics[width=0.45\textwidth]{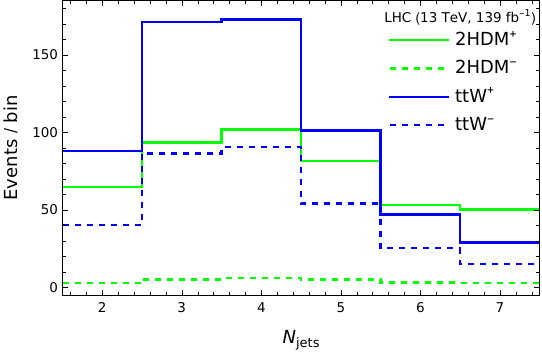}
    \caption{Asymmetry distributions and number of events as function of $N_\text{jets}$ with a scalar $m_H=1000$~GeV, $\rho_{tt}=0.3$ and $\rho_{tu}=0.85$.}
    \label{g2HDM1000A}
\end{center}
\end{figure}

\begin{figure}[t!]
\begin{center}
\includegraphics[width=0.49\textwidth]{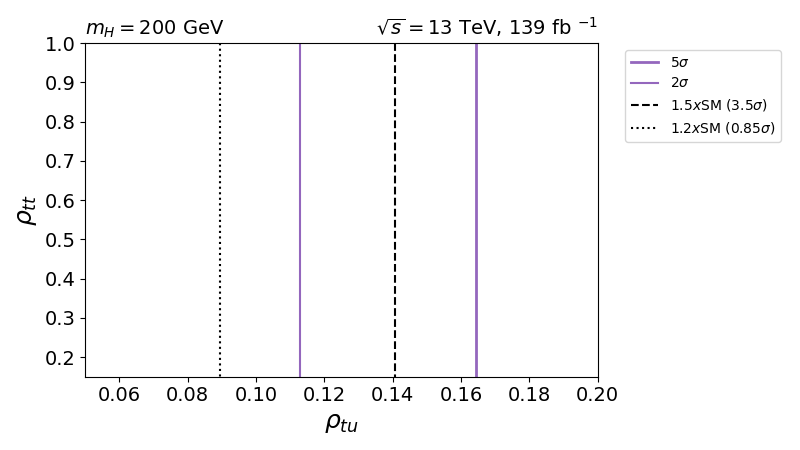}
\includegraphics[width=0.49\textwidth]{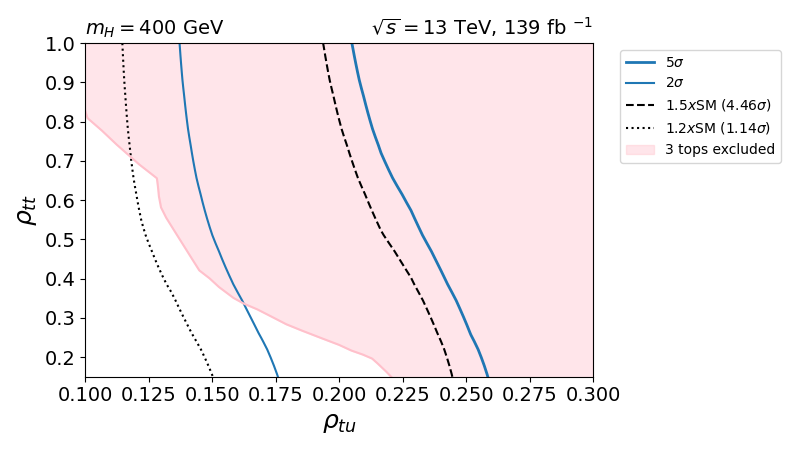}
\vspace{1mm}
\includegraphics[width=0.49\textwidth]{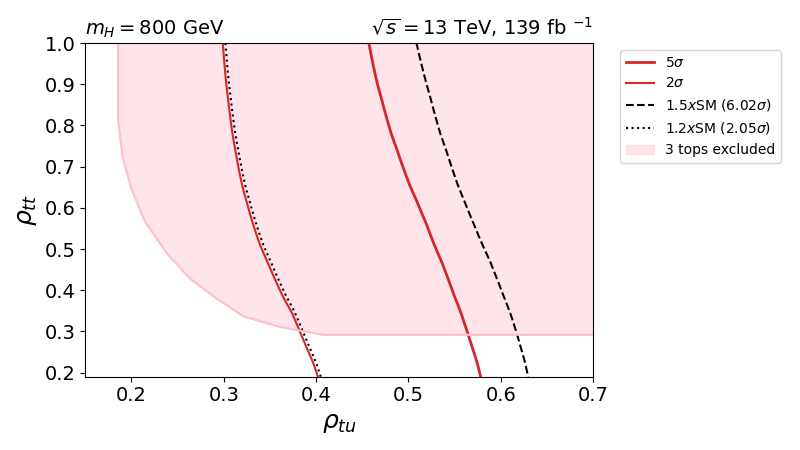}
\includegraphics[width=0.49\textwidth]{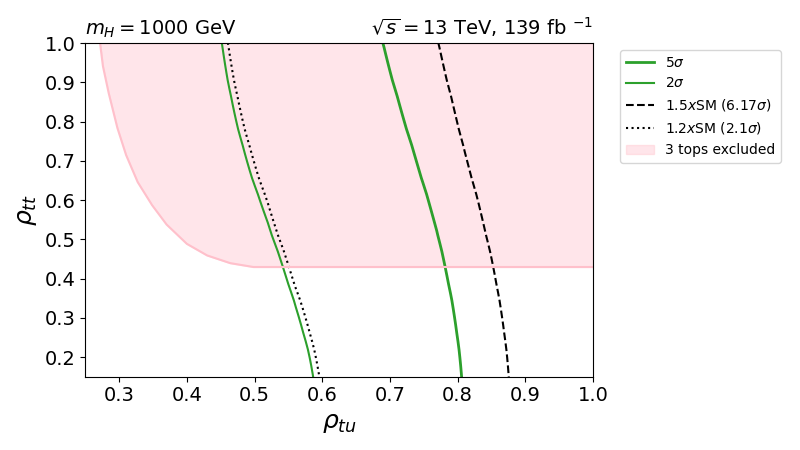}
\caption{Normal Gaussian significance for different masses $m_H=200,\; 400,\; 800,\; 1000$ GeV  at $\sqrt{s}=13$ TeV for $\mathcal{L}=139$ fb$^{-1}$, in the  $\rho_{tu}-\rho_{tt}$ plane. Shaded pink regions are excluded by the currents limits of 3-top production, $\sigma_{3t}<41$~fb~\cite{ATLAS:2023ajo}.}
    \label{g2HDM400B}
\end{center}
\end{figure}

In this part of our analysis we concentrate on three values for $m_H= 200,\; 800,\; 1000$ GeV, though later we also include $m_H=400$ GeV. We verified that taking the lower $m_H$ mass such that $H\to t\bar{t}$ is kinematically allowed ($m_H \gtrsim 350$ GeV) and choosing the largest values for $\rho_{tt}$ and the smallest $\rho_{tu}$ considered, $\rho_{tt}=1$ and $\rho_{tu}=0.1$, we obtain a cross section for four-top-quark production ($t\bar{t}t\bar{t}$; denoted `4-top' in the following) that is below current bounds~\cite{ATLAS:2022rws}. Thus the rest of the parameter space studied in the analysis satisfies the 4-top production bounds. The bounds from three-top-quark production ($t\bar{t}t$+$X$, with $X=q,b,W$; denoted `3-top' in the following), however, tend to be quite strong in constraining the parameter space and we explicitly show that later on in Fig.~\ref{g2HDM400B}. The masses and couplings chosen in Figs.~\ref{g2HDM200A}--\ref{g2HDM1000A} satisfy all experimental bounds.

In Fig.~\ref{g2HDM200A} we display the differential charge asymmetry for $m_H=200$ GeV as a function of the pseudorapidity difference between leptons $\Delta\eta(l,l)$, the azimuthal angle difference between leptons $\Delta\phi(l,l)$, the effective mass ${\rm Eff}_\text{mass}$, the scalar $p_T$ sum $H_{T_\text{jets}}$ and $H_{T_\text{lep}}$ for the jets and leptons, respectively, the lepton-pair invariant mass $m_{{\rm inv}}(l,l)$, the number of $b$-tagged jets $N_b$, and the number of jets $N_{{\rm jets}}$. Notice first of all the clear increment  in the charge asymmetry on all the distributions for the results that include new physics (red histogram) in comparison with only the SM contribution (blue histogram), where the statistical error bands are displayed in light red and light blue, respectively. This increment of roughly a factor $\sim 1.5$  clearly shows the  power that  this observable has in this particular model for distinguishing new physics from the SM contribution. Some of the charge asymmetry distributions are quite flat as depicted for $\Delta\eta(l,l)$, $\Delta\Phi(l,l)$, $H_{T_\text{lep}}$, and $N_b$, and this flatness remains for higher $m_H$ masses, thus we do not show these distributions for $m_H=800$ GeV and $m_H=1000$ GeV. Others like the ones for the effective mass, $H_{T_\text{jets}}$, and $N_{{\rm jets}}$ show a clear trend and for $m_H=200$ GeV they peak toward lower values of these respective quantities, whereas, on the other hand, for $m_{{\rm inv}}(l,l)$ there is a slight increase towards larger values. Lastly notice that in the histogram in Fig.~\ref{g2HDM200events} of number of events versus $N_{{\rm jets}}$, there are more events at smaller $N_{{\rm jets}}$ for $m_H=200$ GeV. When we analyze Fig.~\ref{g2HDM400A}, corresponding to $m_H=800$ GeV, we see that now all of the distributions are starting to increase at larger values of the quantities they are plot against with. This shift becomes even more evident in Fig~\ref{g2HDM1000A} for $m_H=1000$ GeV, in which all distributions peak toward the largest values that the histograms are done against with and charge asymmetries as large as a factor of 2.26 with respect to the SM contribution can be obtained, for example in $N_{{\rm jets}}\geq 7$ bin.

Having obtained the result for a choice of couplings, one can easily get the differential charge asymmetry for a different choice of couplings by rescaling appropriately. Performing  a binned $\chi^{2}$-fit with 2 degrees of freedom for the red and blue histograms of the $N_{{\rm jets}}$ distribution, and taking into account the statistical errors, we calculate a Gaussian significance, for a fixed value of $m_H=200,\; 400,\; 800,\; 1000$ GeV , in the $\rho_{tu}$ vs $\rho_{tt}$ plane, as shown in Fig.~\ref{g2HDM400B}. The thinner and broader solid lines, shown for each mass, correspond to 2$\sigma$ and 5$\sigma$ significance. They can loosely be interpreted as new-physics hint and discovery threshold, respectively, and should be used as reference. The dashed and dotted lines correspond to signal cross sections that are 50 and 20 percent respectively increased with respect to the SM $t\bar{t}W$ signal, as allowed by the latest theoretical predictions and experimental measurements~\cite{ATLAS:2023gon}.  The different colors correspond to the different $m_H$ masses considered. Also shown is the 3-top search constraints as a shaded pink area, which clearly puts a strong constraint for $m_H \gtrsim 350$ GeV, even at $m_H=1$ TeV. As is clear from the figure, for large values of $\rho_{tt}$, the branching ratio $BR(H\to \bar{t}t)$ saturates to one, leading to a signal that only depends on $\rho_{tu}$. As $\rho_{tt}$ decreases, the contribution from the $tt\bar{t}$ channel decreases and larger values of $\rho_{tu}$ are necessary to compensate by increasing the production cross sections. The necessary increment in $\rho_{tu}$ grows with larger $m_H$. Note also that for small $m_H$ one could see a hint only for the largest allowed signal cross sections and that for those cross-sections values, as $m_H$ increases one would obtain significances above discovery. On the other hand, for the smaller increment considered on the signal cross section one could  barely reach a hint at $m_H=1000$ GeV.

An interesting remark is that, recalling the CP independence of these results, the point in the parameter space $\rho_{tt}\approx 0.7$, $\rho_{tu}\approx 0.23$ for a CP-odd scalar with $m_A=400$ GeV for which we would obtain a significance of $\sim 5\sigma$ could also be compatible and responsible for explaining  another somewhat recent collider anomaly observed in the  ditop search performed by the CMS collaboration~\cite{CMS:2019pzc} at $\sqrt{s}=13$ TeV and a luminosity of $\mathcal{L} = 35.9$ fb$^{-1}$, with top quarks decaying into single and dilepton final states. It was shown that  deviations from the SM behavior arise at the $3.5\sigma$ level locally  ($1.9\sigma$ after the look-elsewhere effect) which can be accounted for by a pseudoscalar with a mass around 400 GeV that couples at least to top quarks at tree-level~\cite{Arganda:2021yms}. Sadly the latest 3-top searches have ruled out this possibility, as can be seen in Fig.~\ref{g2HDM400B} for that mass and choice of couplings, though smaller values of $\rho_{tu}$ are still compatible with this collider anomaly, $\rho_{tu}\lesssim 0.125$.

\begin{figure}[t!]
\begin{center}
\includegraphics[width=0.45\textwidth]{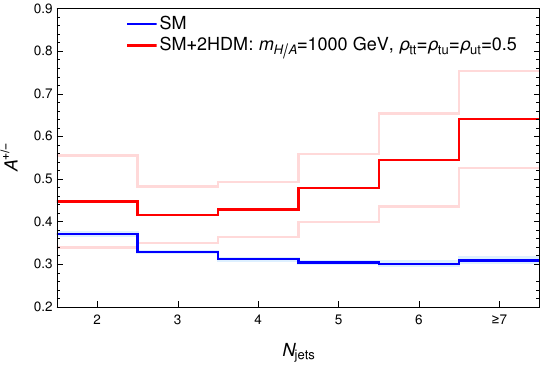}
\caption{Asymmetry with both CP even and CP odd a scalars $M_H=M_A=1000$~GeV and $\rho_{tt}=\rho_{tu}=0.5$.}
\label{g2HDMAH1000}
\end{center}
\end{figure}

In the g2HDM the squared mass difference between $A$ and $H$ is proportional to a quartic coupling and the SM Higgs vev squared. Thus, unless such quartic coupling takes values much greater than one, one would naturally expect both states to have similar masses. In that case, both states would contribute in the diagrams of Fig.~\ref{g2HDMdiags}. As an illustrative example,  we show in Fig.~\ref{g2HDMAH1000} the charge asymmetry as a function of $N_{{\rm jets}}$ for $M_H=M_A=1000$~GeV and $\rho_{tt}=\rho_{tu}=0.5$, which is barely excluded by 3-top searches.  We checked that there is no interference between the CP-even and CP-odd contributions and, thus,  we roughly obtain double the amount of asymmetry as we would have for either $A$ or $H$ alone. The same applies to the rest of the distributions analyzed.  

\subsection{Effective theory with four-quark operators}
\label{EFT}

We consider the case in which the new BSM states are too heavy to be produced at the LHC, such that their effects can be described by an effective low energy theory, with higher dimensional operators added to the SM~\cite{Buchmuller:1985jz}. Since we want to study the charge asymmetry, we consider only four-quark operators and we work with the following basis:
\begin{align}
& {\cal Q}^{(1,1)}_q = (\bar Q_L \gamma^\mu Q_L)^2 \ , \nonumber\\
& {\cal Q}^{(1,3)}_q = (\bar Q_L \gamma^\mu t^j Q_L)^2 \ , \nonumber\\
& {\cal Q}^{(1)}_u = (\bar u_R \gamma^\mu u_R)^2 \ , \nonumber\\
& {\cal Q}^{(1)}_{qu} = (\bar Q_L \gamma^\mu Q_L)(\bar u_R \gamma_\mu u_R) \ , \label{eq-Q6}
\end{align}
where $t^j$ are the SU(2)$_L$ generators and generation indices are implicit . There are also operators with color octet currents: ${\cal Q}^{(8,1)}_q$, ${\cal Q}^{(8,3)}_q$ ,${\cal Q}^{(8)}_u$, ${\cal Q}^{(8)}_{qu}$, that can be obtained from Eq.~(\ref{eq-Q6}) by inserting color generators in the quark bilinears. Other four-quark operators can be obtained by Fierz transformations. We assume that the Wilson coefficient of operators with a top-antitop pair and bosonic fields are very small and we will not consider them in this section. As an example, operators with gluons would contribute to top-antitop production with initial bosonic states that are charge symmetric, thus diluting the charge asymmetry by increasing its denominator.

For our analysis, instead of considering the most general set of operators that would lead to a multidimensional space of Wilson coefficients difficult to interpret, we focus for simplicity on the case with just the following two operators:
\begin{equation}
{\cal L}_{\rm eff} = [C^{(1)}_{u}]^{3131}{\cal Q}^{(1),3131}_{u} + [C^{(1)}_{qu}]^{3131}{\cal Q}^{(1),3131}_{qu} \ ,
\label{eq-4q}
\end{equation}
where the superindices number generations of the quarks involved in the operators. We are agnostic about the high energy physics generating ${\cal L}_{\rm eff}$, we just choose two operators involving the first and third generations, to study their interplay in the charge asymmetry. To leading order, operators involving charm and top quarks will dilute the charge asymmetry, see the introduction of Section~\ref{BSM}. In the following we trade $[C^{(1)}_{u}]^{3131}\to C_u$ and $[C^{(1)}_{qu}]^{3131}\to C_{qu}$, to simplify the notation, and we will refer to the model as 4qFCNC.

\begin{figure}[t!]
\begin{center}
\begin{tabular}{cc}
\includegraphics[width=0.45\textwidth]{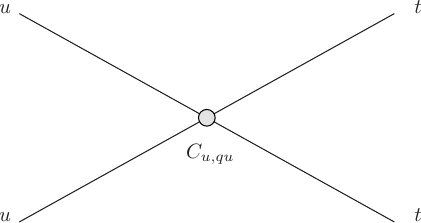}
\hspace{1cm} &
\includegraphics[width=0.45\textwidth]{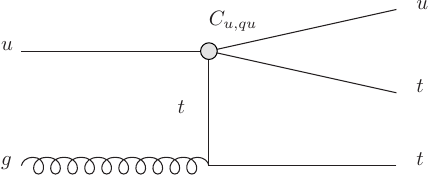} \\
(a)\hspace{1cm} & (b)
\end{tabular}
\caption{Leading order contributions to $tt$ production in the effective theory with four-quark operators. These diagrams and their conjugates contribute to the 2SSl plus jets final state and the charge asymmetries.}
\label{fd-4q}
\end{center}
\end{figure}

The operators from Eq.~(\ref{eq-4q}) contribute to the charge asymmetry with the Feynman diagrams of Fig.~\ref{fd-4q}, where we show the production of $tt$ and $ttj$, similar diagrams can be obtained for $\bar t\bar t$ production. Since the later process is initiated by $\bar u$ in the initial state, it is suppressed by PDFs at the LHC compared with the former one. 

We have checked that for the Wilson coefficients considered, the bounds from 4-top production are satisfied. Given that the amplitude is proportional to $C_i^2$, these contributions are suppressed, for the couplings considered the cross section is increased by less than 5\% compared with the SM one. The contribution to 3-top production is also negligible and below the current bounds~\cite{ATLAS:2023ajo}.

\begin{figure}[t!]
\begin{center}
\includegraphics[width=0.45\textwidth]{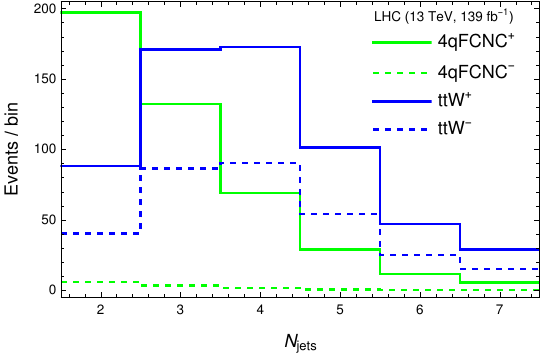}
\caption{Number of events as function of $N_\text{jets}$, for the effective theory with four-quark operators,
with $C_u=0.062/{\rm TeV}^2$ and $C_{qu}=0$. 
}
\label{yield-4q}
\end{center}
\end{figure}

In Fig.~\ref{yield-4q} we show the number of events as functions $N_\text{jets}$, for the SM background and signal, classifying them in terms of the total charge. For the signal we show the result for $C_u=0.062/{\rm TeV}^2$ and $C_{qu}=0$, that roughly gives 50\% of the total number of events of the SM. Note that this value of $C_u$, for a coupling of order 1, points to an effective scale of 4 TeV. For the SM the positive charge is approximately twice of the negative one, with the maximum for 3 and 4 jets. For the signal, the number of negative charge events is very small, and the positive one is peaked towards small number of jets, rapidly decreasing as $N_\text{jets}$ increases.

\begin{figure}[t!]
\begin{center}
\includegraphics[width=0.45\textwidth]{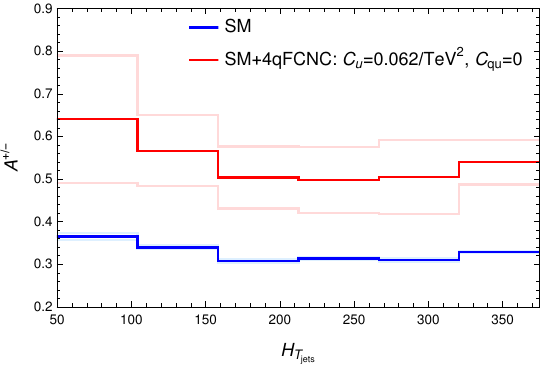}
\includegraphics[width=0.45\textwidth]{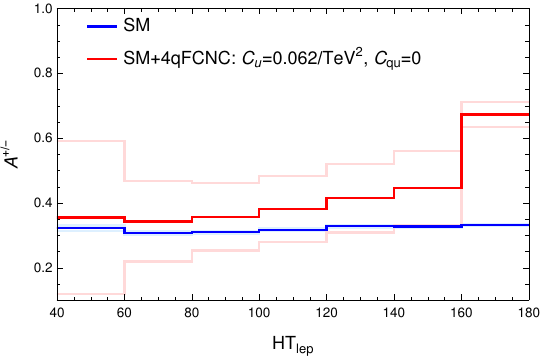}
\vspace{1mm}
\includegraphics[width=0.45\textwidth]{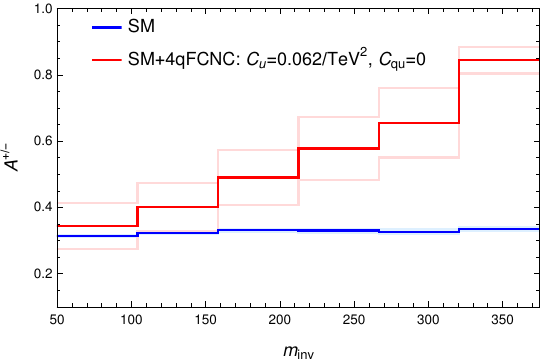}
\includegraphics[width=0.45\textwidth]{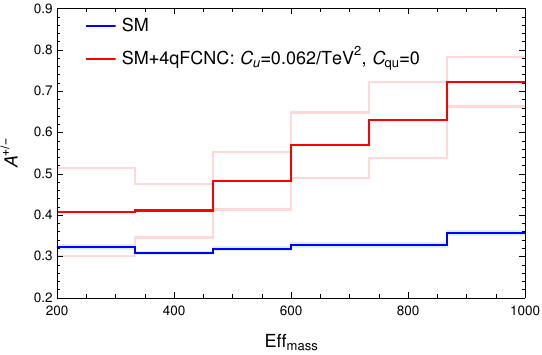}
\vspace{1mm}
\includegraphics[width=0.45\textwidth]{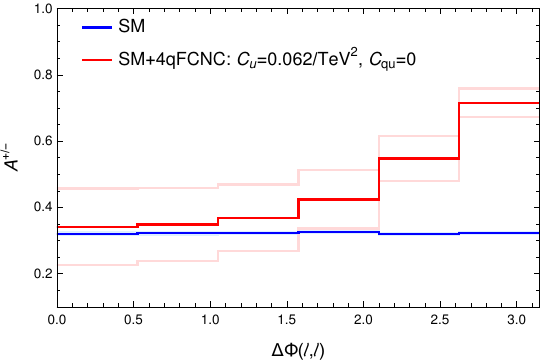}
\includegraphics[width=0.45\textwidth]{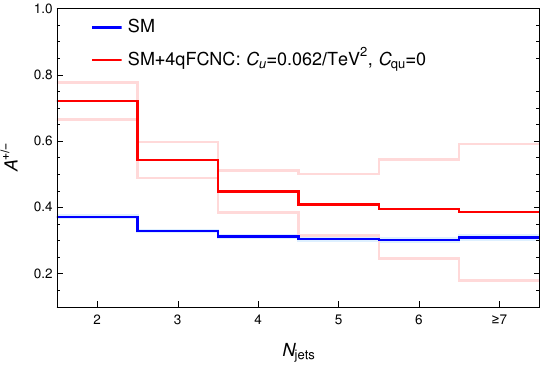}
\caption{Asymmetry distributions in the effective theory with four-quark operators
, with $C_u=0.062/{\rm TeV}^2$ and $C_{qu}=0$.}
\label{dist-4q}
\end{center}
\end{figure}

Generating signal events with the same pipeline as in Section~\ref{g2HDM}, in Fig.~\ref{dist-4q} we show the distributions of the charge asymmetry in terms of the same variables as in Section~\ref{g2HDM}, for $\sqrt{s}=13$ TeV and $\mathcal{L}=139$ fb$^{-1}$, with the same Wilson coefficients: $C_u=0.062/{\rm TeV}^2$ and $C_{qu}=0$. The total charge asymmetry is significantly increased with respect to the SM, the distributions also have different shapes compared with the SM, as well as with the case of a pseudo/scalar field. The distribution of $H_{T_\text{jets}}$ is peaked towards small $H_{T_\text{jets}}$, like a light scalar, although in that case it always decreases as $H_{T_\text{jets}}$ increases, whereas for a heavy scalar the peak is towards large $H_{T_\text{jets}}$. The distribution of $H_{T_\text{lep}}$ is strongly peaked towards large $H_{T_\text{lep}}$, like a heavy scalar, although in that case the peak is less prominent, whereas for a light scalar it is rather flat. The distribution of $m_\text{inv}$ is also peaked towards large $m_\text{inv}$, as for a heavy scalar, again in that case the peak is less prominent, and a light scalar is again rather flat. The distribution of Eff$_\text{mass}$ is also peaked towards large Eff$_\text{mass}$, as for a heavy scalar, but in that case the peak is somewhat higher, whereas for a light scalar it is peaked towards small values. The distribution of $\Delta\phi(l,l)$ is peaked towards large $\Delta\phi(l,l)$, as for a light scalar, but in that case the peak is somewhat lower. Finally, the distribution of $N_\text{jets}$ is peaked towards small $N_\text{jets}$, as for a light scalar, but in that case the peak is somewhat lower, whereas for a heavy scalar it is peaked towards large $N_\text{jets}$. 

\begin{figure}[t!]
\begin{center}
\includegraphics[width=0.55\textwidth]{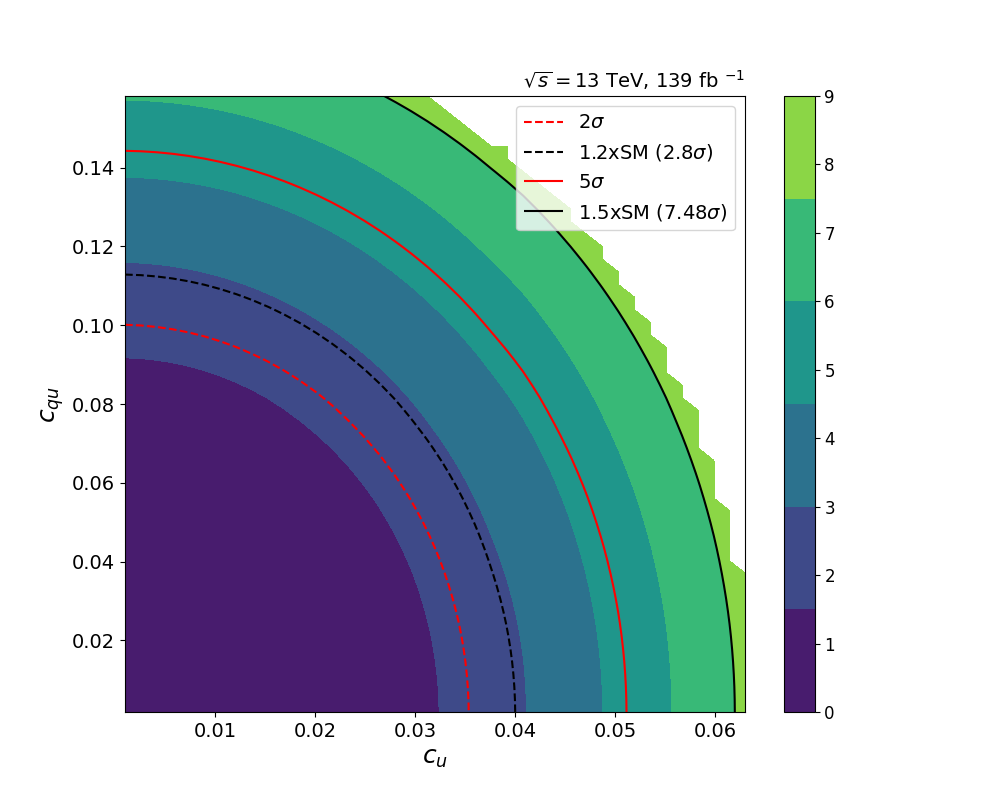}
\caption{Normal Gaussian significance curves of 2$\sigma$ and 5$\sigma$, as well as cross sections 20 and 50\% larger than the SM alone, for the effective theory with four-quark operators of Eq.~\ref{eq-4q}.
}
\label{2d-4q}
\end{center}
\end{figure}

Proceeding similarly to Section~\ref{g2HDM}, in Fig.~\ref{2d-4q} we show curves of 2$\sigma$ and 5$\sigma$, as well as corrections to the cross sections of 2SS$l$ of 20 and 50\%, in the plane of couplings. Matching $C=g_{\rm NP}^2/m_{\rm NP}^2$, for $g_{\rm NP}=1$, the curves of 2$\sigma$ and 5$\sigma$ point to scales $m_{\rm NP}=2-2.4$~TeV ($1.6-1.8$~TeV) when only $C_u$ ($C_{qu}$) is active. As an example, $C_u$ can be associated with a $Z'$ resonance coupled to Right-handed up-type quarks, integrated-out. The effective operators analyzed in this section cannot be associated to a heavy scalar having been integrated out; however, by comparing this case with a scalar of 1 TeV, a larger effect is obtained due to the three polarizations of the massive spin one particle.

\subsection{RPV Supersymmetry}
\label{RPV-SUSY}
The lack of supersymmetric signatures at the LHC leads to the necessity of relaxing some of the usual assumptions made about these types of theories. It has been known for a long time that the usual assumption of an $R$-parity conserving theory, though preventing proton decay and stabilizing the lightest supersymmetric particle, can be eased allowing $R$-parity violating couplings~\cite{Barbier:2004ez} that do not lead to proton decay but make the LSP unstable, decaying to SM particles. On the other hand, but related to the lack of SUSY signatures at the LHC, light colored SUSY particles are increasingly constrained, pushed to higher masses. Thus it becomes natural to consider scenarios in which colored particles are heavy and decoupled at the energies probed at the LHC. Is in that spirit that we work within the framework of the MSSM with non-conserved $R$-parity, so that supersymmetric particles do not have to be produced in pairs, they are not stable, and their decay chains do not end up in the lightest supersymmetric particle (LSP) but in SM particles~\cite{Dreiner:2023bvs}. The MSSM interactions responsible for $R$-parity violation come from the superpotential, that reads as~\cite{Barbier:2004ez}:
\begin{equation}
W_\text{RPV} = \mu_i L_i H_u + \frac{1}{2} \lambda_{ijk} L_i L_j E_k^c + \lambda^\prime_{ijk} L_i Q_j D_k^c + \frac{1}{2} \lambda^{\prime\prime}_{ijk} U_i^c D_j^c D_k^c \,,
\end{equation}
where $Q_j$ is the $SU(2)_L$ quark isodoublet superfield, $U_i^c$ is the right-handed up-type quark isosinglet superfield and $D_j^c$ is the down-type one, while the $SU(2)_L$ lepton isodoublet and isosinglet superfields are $L_i$ and $E_i^c$, respectively, and $H_u$ is the up-type Higgs superfield responsible for the the up-type quarks masses. In order to avoid the proton decay, $\lambda^{\prime\prime}$ = 0 is assumed.

\begin{figure}[t!]
\begin{center}
\begin{tabular}{cc}
\includegraphics[width=0.35\textwidth]{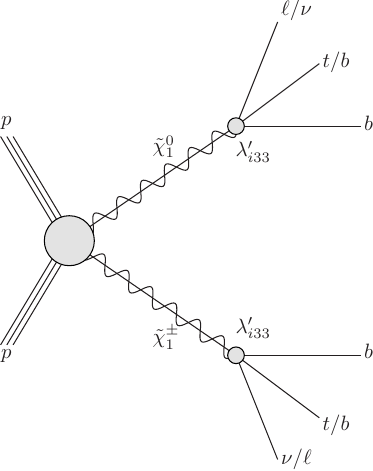}\hspace{2cm} &
\includegraphics[width=0.35\textwidth]{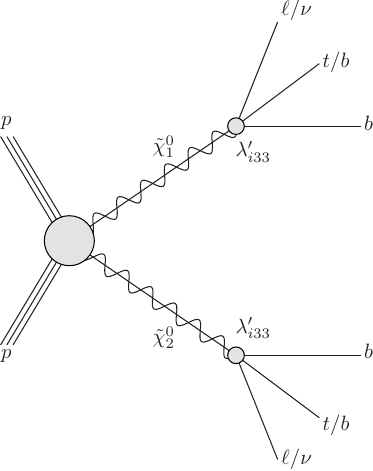}\\
(a)\hspace{2cm}  & (b) 
\end{tabular}
\caption{Diagrams of RPV-SUSY electroweakino production at the LHC that contribute to the 2SSl plus jets final state and the charge asymmetries: (a) $\tilde{\chi}_1^0$-$\tilde{\chi}_1^\pm$ production; (b) $\tilde{\chi}_1^0$-$\tilde{\chi}_2^0$ production.}\label{RPV-SUSY_diags}
\end{center}
\end{figure}

As we are interested in final states with leptons and quarks, the new physics source of charge asymmetries within this RPV MSSM framework are generated from $\lambda^{\prime}$ interaction terms, which are obtained through the expansion of the $\lambda^{\prime}$ superpotential term as function of fermion and sfermion (scalar fermions) fields as follows~\cite{Barbier:2004ez,Deshpande:2016yrv,Das:2017kfo}:
\begin{equation}
{\cal L} = \lambda^\prime_{ijk} \left[ \tilde \nu_L^i \bar d_R^k d_L^j + \tilde d_L^j \bar d_R^k \nu_L^i + \tilde d_R^{k\ast} \bar \nu_L^{ci} d_L^j - \tilde \ell_L^i \bar d_R^k u_L^j - \tilde u_L^j \bar d_R^k \ell_L^i - \tilde d_R^{k\ast} \bar \ell_L^{ci} u_L^j \right] \,.
\end{equation}
Here tildes denote sfermions and $c$ corresponds to charge-conjugated fields.

We work within two different RPV-SUSY scenarios, depending on the values of the gaugino/higgisno mass parameters:
\begin{enumerate}
\item Higgsino-like scenario: Only the $\mu$ parameter is of electroweak order, achieving a spectrum with two light higgsino-like neutralinos and a light higgsino-like chargino, while the rest of the SUSY spectrum is heavy.
\item Wino-like scenario: $M_1\sim M_2$ of order electroweak, being $M_1$ and $M_2$ the bino and the wino masses, respectively, while the rest of the SUSY mass parameters are heavy. In this case we have two light gaugino-like neutralinos and a light wino-like chargino.
\end{enumerate}
Taking these spectra into account, we consider a simplified model with electroweakino production (higgsino-like or wino-like), i.e. the type of diagrams of Fig.~\ref{RPV-SUSY_diags}. The masses of $\tilde \chi^{\pm}_1$, $\tilde \chi^0_1$, and $\tilde \chi^0_2$ are degenerate, considering the range from 200 GeV to 800 GeV~\footnote{As a caveat, in relation to this choice of mass range for winos and higgsinos, it should be clarified that values below about 600 GeV may be excluded by current searches~\cite{ATLAS:2023tlp}. However, we prefer to work also with these masses to study the potential of charge asymmetry, in the most favorable scenarios, as an observable to distinguish between new physics and SM.}. We set the rest of the SUSY particle masses to few TeV order. The RPV couplings we consider are $\lambda^\prime_{i23}$ ($i =$ 2, 3) (motivated by the $B$ anomalies~\cite{Deshpande:2016yrv,Das:2017kfo,Earl:2018snx,Hu:2019ahp,Hu:2020yvs,Heinemeyer:2021opc,Zheng:2021wnu,Choudhury:2023lbp}) which give rise to the following neutralino/chargino decay modes:
\begin{eqnarray}
&& \tilde \chi^{\pm}_1 \to b b \ell \,, b t \nu \,, \\
&& \tilde \chi^0_1/\tilde \chi^0_2 \to b t \ell \,, b b \nu \,.
\end{eqnarray}
We set $\lambda^\prime_{223}$ $=$ $\lambda^\prime_{333}$; the exact value is not important as long as it is $\gtrsim$ $10^{-4}$ so that the decay is prompt. This is the choice made in the most recent ATLAS search for this model~\cite{ATLAS:2023tlp}; as stated before, we are considering illustrative models where a differential charge asymmetry could potentially enhance the current sensitivity. It is important to note that plugging the coupling with electrons should not change anything with respect to the charge asymmetry, since the asymmetry is dictated by the flavor of the initial state partons, which in the case of neutralino/chargino production is $t \bar t W$-like.

\begin{figure}[t!]
\begin{center}
\includegraphics[width=0.45\textwidth]{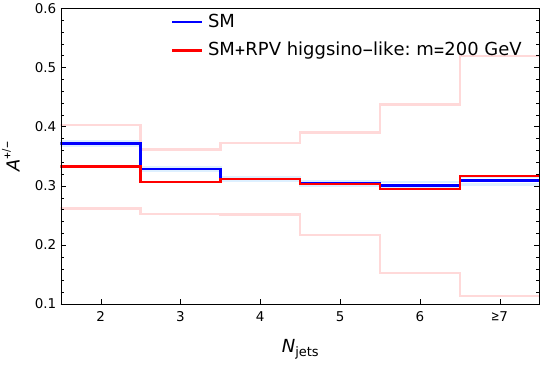}\hspace{15mm}
\includegraphics[width=0.45\textwidth]{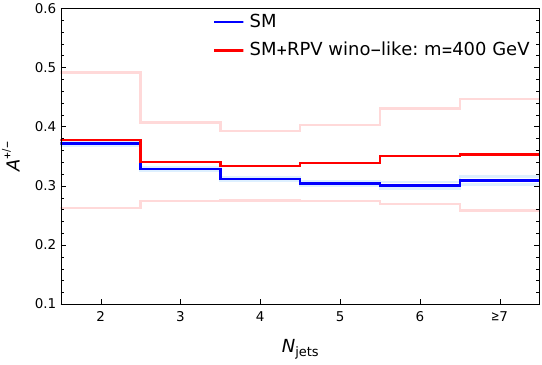}
\caption{Differential charge asymmetry as a function of $N_\text{jets}$ in RPV SUSY. Left panel: higgsino-like neutralinos and charginos with masses of 200 GeV. Right panel: wino-like neutralinos and charginos with masses of 400 GeV.}
\label{dist-Asym-Njets_RPV-SUSY}
\end{center}
\end{figure}

As in the previous sections, we used {\sc MadGraph5\_aMC@NLO}+{\sc Pythia}+{\sc Delphes} to generate sample events, but without including spin correlations, since their effects in this scenario are expected to be negligible. In Fig.~\ref{dist-Asym-Njets_RPV-SUSY} we show the results of the distributions of the differential charge asymmetries as a function of the number of jets $N_\text{jets}$ for the most favorable scenarios: 200-GeV higgsino-like neutralinos and charginos (left panel) and 400-GeV wino-like neutralinos and charginos (right panel). We take these two cases because they are the most similar (quite roughly) to having 1.5 $\times$ $\sigma_{t \bar t W}$. In fact, masses lower than 400 GeV in the wino-like case are discarded with that criterion, because they give rise to much more new physics events not compatible with $t \bar t W$ cross-section measurement at the LHC. With this same criterion, it is not superfluous to clarify that masses higher than 200 GeV in the higgsino-like scenario and masses higher than 400 GeV in the wino-like scenario are not only not ruled out by dedicated searches~\cite{ATLAS:2023tlp}, but are obviously even less sensitive than the cases defined as favorable, since the higher the mass, the lower the number of new physics events, so one logically obtains less sensitivity.

One can see clearly from both panels of Fig.~\ref{dist-Asym-Njets_RPV-SUSY} that the results, in the most favorable RPV-SUSY scenarios considered, are indistinguishable from those of the SM, but well within the error bands. The significance, obtained from the $\chi^2$ of only one degree of freedom (the neutralino/chargino mass), for the higgsino-like case with 200 GeV mass is as low as 0.007, and for the wino-like case with 400 GeV mass is 0.51: this difference is partly due to the fact that the higgsino-like case is actually 1.55 $\times$ $\sigma_{t \bar t W}$ while the wino-like is 1.63 $\times$ $\sigma_{t \bar t W}$. In line with the comment in the previous paragraph, scenarios with higher masses are not probed since they contribute even less to the charge asymmetry. It should also be noted that the charge asymmetry as a function of the other observables considered is not sensitive either and, therefore, we use $N_\text{jets}$ as an illustrative example.

We can then conclude from these results that the charge asymmetry is not a good observable to distinguish new physics in these RPV-SUSY scenarios in the parameter space considered along this work. We would also like to comment that considering the effective operator from integrating out the neutralino/chargino sector this would lead to higher dimensional operators of dimension 11, which obviously are strongly suppressed and cannot provide large values of charge asymmetries.

\section{Prospects for HL-LHC}
\label{sec-hllhc}

It is interesting to study the prospects for the charge asymmetry observable in the final states considered at the high luminosity (HL)-LHC. To get a rough estimate, we study in this section the asymmetry at $\sqrt{s}=13$ TeV, with a luminosity $\mathcal{L}=3000$ fb$^{-1}$, which in practice ends up reducing the statistical uncertainty. Even though for the HL-LHC the energy will likely be running at 14 TeV, we expect our analysis to provide a conservative estimate.

In addition to the default inclusive 1 $b$-jet selection that we have considered so far at  $\mathcal{L}=139$ fb$^{-1}$, we shall consider as well a selection with different cuts in the number of $b$-jets: exclusive 1 and 2 $b$-jet selections, an inclusive 3 $b$-jet selection, and finally the statistical combination of the latter three, in order to investigate the sensitivities one could reach in different signal regions.

\begin{figure}[b!]
\begin{center}
\includegraphics[width=0.45\textwidth]{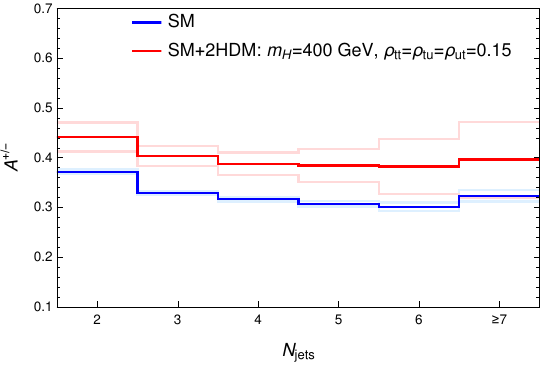}
\includegraphics[width=0.45\textwidth]{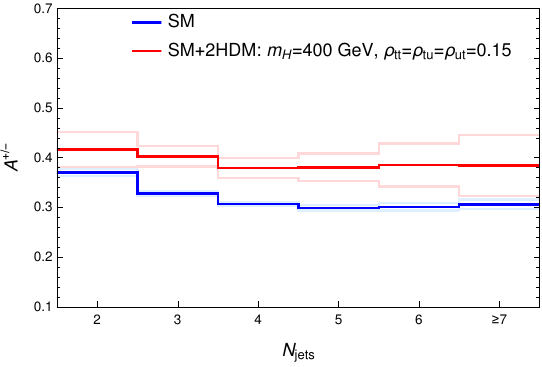}
\vspace{1mm}
\includegraphics[width=0.45\textwidth]{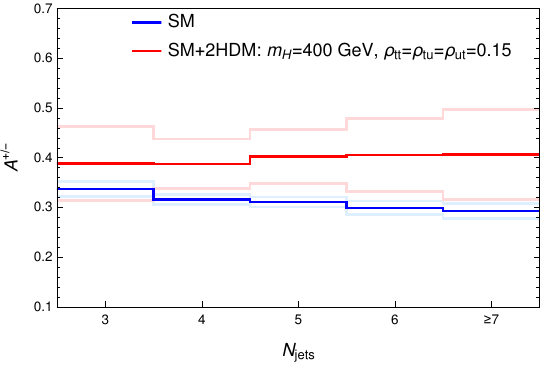}
\includegraphics[width=0.45\textwidth]{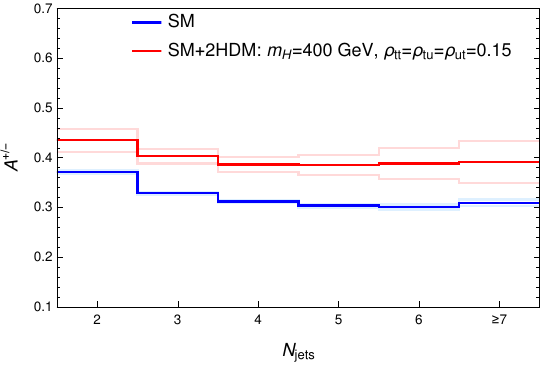}
\caption{Asymmetry vs $N_\text{jets}$, in the 2HDM with a scalar of 400 GeV, at the HL-LHC. On the top left we show the exclusive 1 $b$-jet selection, on the top right the exclusive 2 $b$-jet selection, on the bottom left the inclusive 3 $b$-jet one and on the bottom right the 1 $b$-jet inclusive.}
\label{HLLHC-2HDMNj}
\end{center}
\end{figure}

\begin{figure}[b!]
\begin{center}
\includegraphics[width=0.45\textwidth]{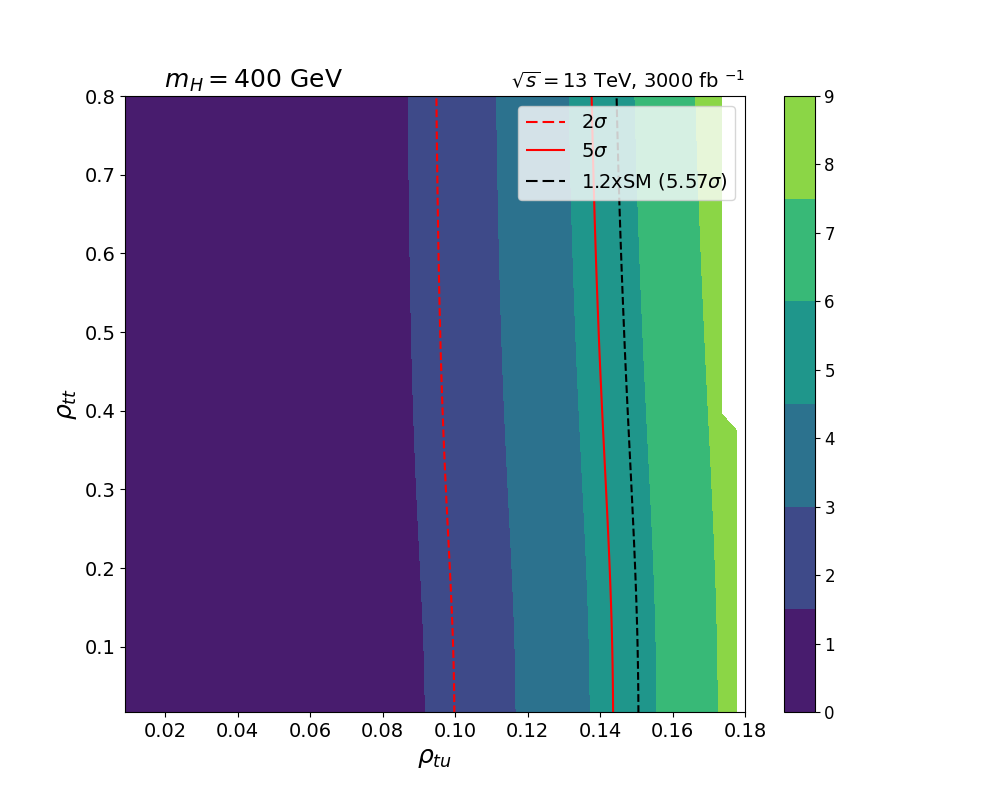}
\includegraphics[width=0.45\textwidth]{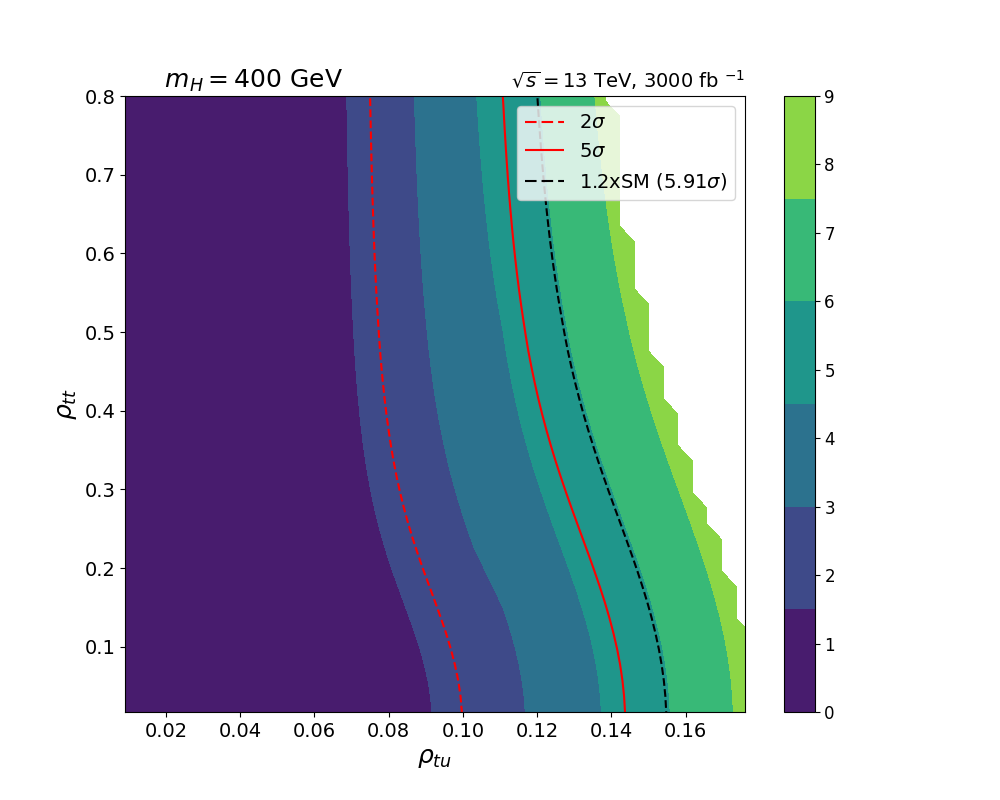}
\vspace{1mm}
\includegraphics[width=0.45\textwidth]{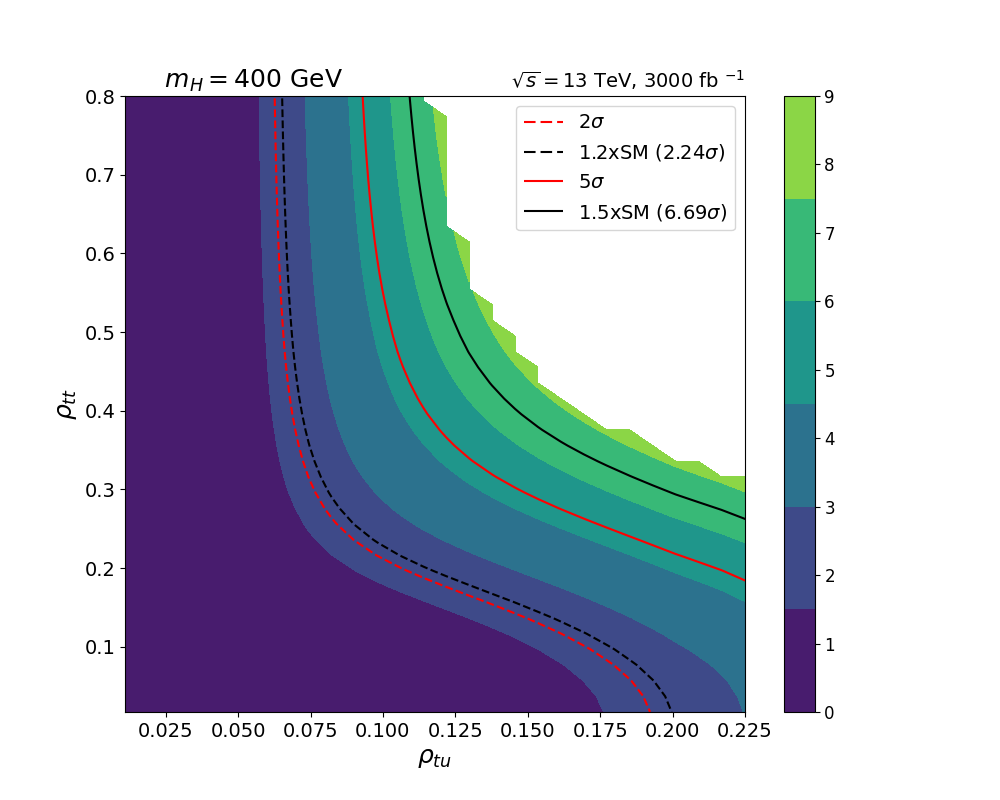}
\includegraphics[width=0.45\textwidth]{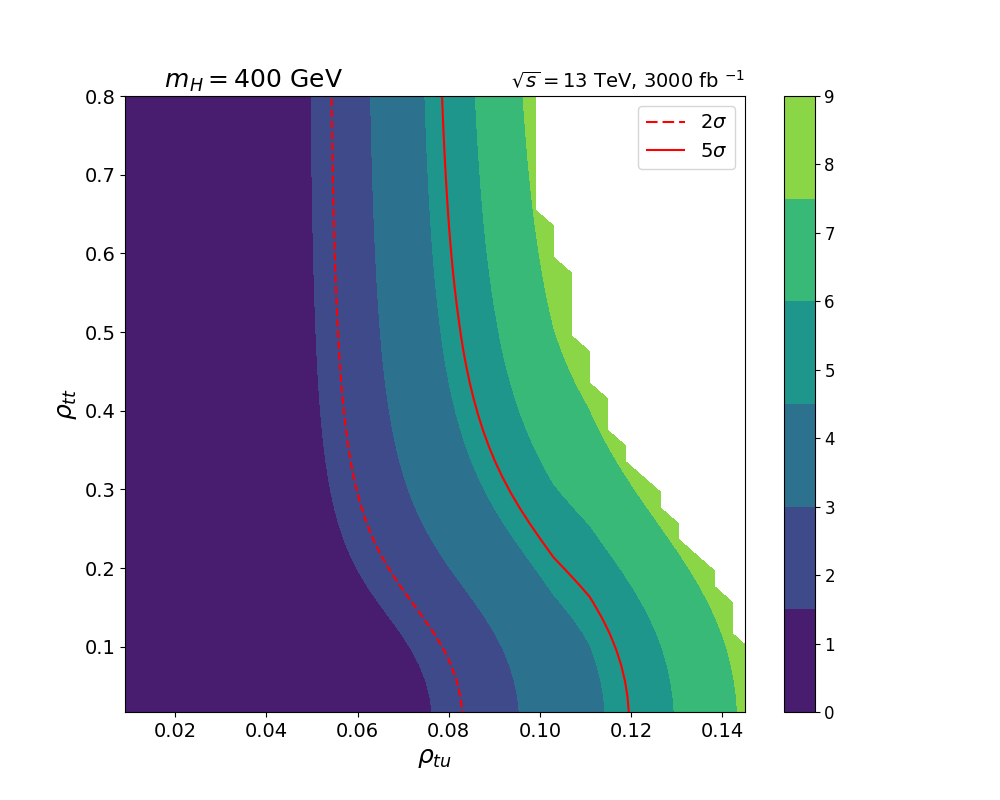}
\vspace{1mm}
\includegraphics[width=0.45\textwidth]{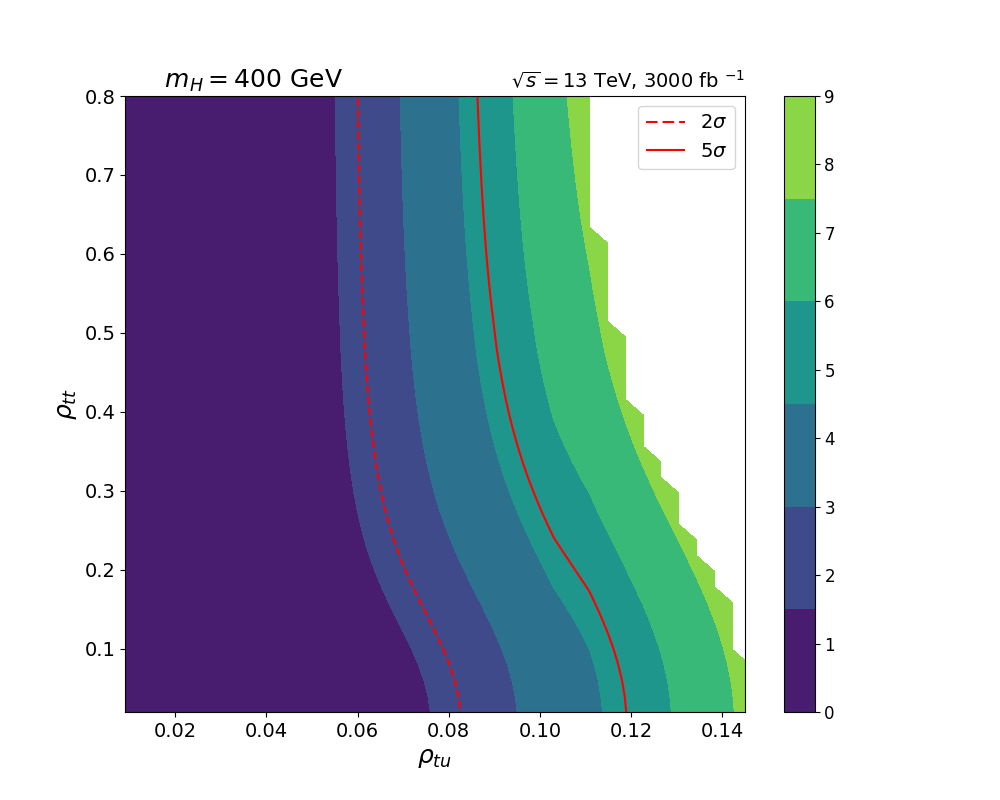}
\caption{Normal Gaussian significance curves for the 2HDM, with a scalar of 400 GeV, for the HL-LHC. In the first line, we show on the left the exclusive 1 $b$-jet selection and on the right the exclusive 2 $b$-jet selection, in the second line we show on the left the inclusive 3 $b$-jet one and on the right the combined analysis, in the third line the inclusive 1 $b$-jet selection.}
\label{HLLHC-2HDMb}
\end{center}
\end{figure}

In the case of the g2HDM we focus on a scalar of 400 GeV that roughly gives 20\% of the total number of events of the SM, showing the asymmetry as functions of $N_\text{jets}$ in Fig.~\ref{HLLHC-2HDMNj}, for the different selections in the number of $b$-jets. It is possible to observe the reduction of the uncertainty, as expected for a larger luminosity. Furthermore, all the different selections are above the SM, remaining quite flat as the number of $b$-jets increases. The combined analysis of the three alternative $b$-jets selections is, as expected, more sensitive than the inclusive 1 $b$-jet selection, in particular at larger values of $\rho_{tt}$. In Fig.~\ref{HLLHC-2HDMb}, we show significance curves for the same $b$-jets selections, and as can be appreciated, in all the cases it is possible to test discovery for points with a cross section 50\% larger than the SM prediction, in contrast with the results we were finding for the LHC at $\mathcal{L}=139$ fb$^{-1}$, as displayed in Fig.~\ref{g2HDM400B}. 

\begin{figure}[b!]
\begin{center}
\includegraphics[width=0.45\textwidth]{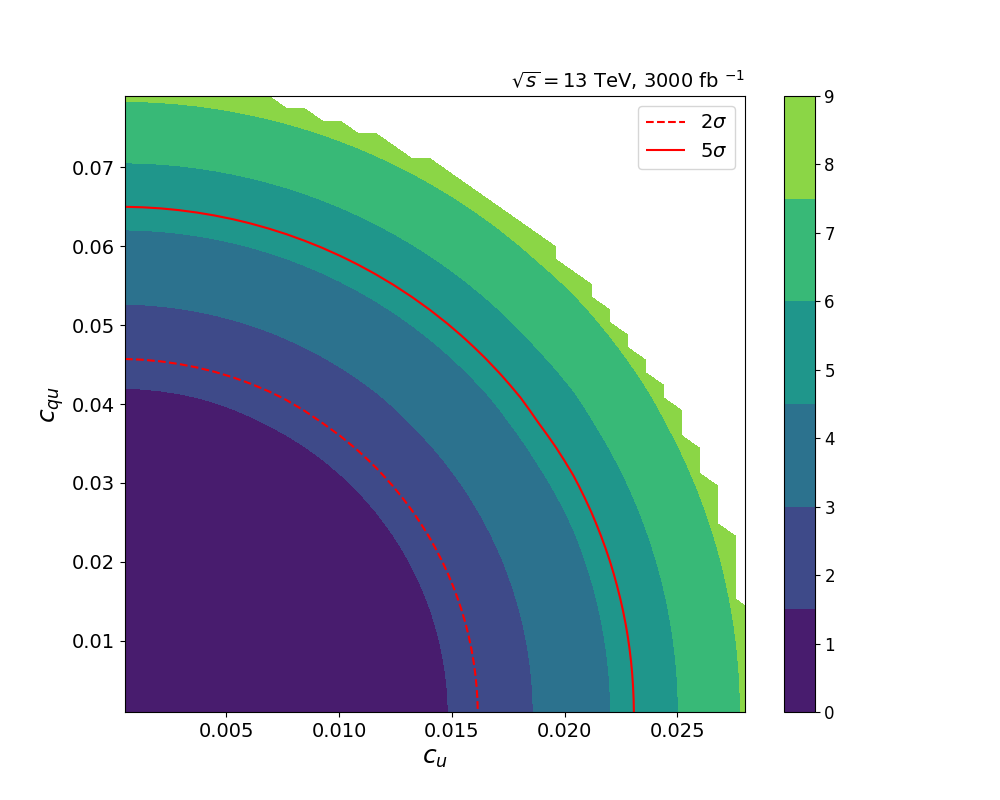}
\caption{Normal Gaussian significance curves for the effective theory with 4-quark operators of Eq.~\ref{eq-4q}, for the HL-LHC, with an exclusive 1 $b$-jet selection.}
\label{HLLHC-4q}
\end{center}
\end{figure}

For the effective theory with four-quark operators, we show in Fig.~\ref{HLLHC-4q} the significance curves on the plane of Wilson coefficients $C_u$ and $C_{qu}$, for the exclusive 1 $b$-jet selection. Other $b$-jet selections are less sensitive than this one, and the improvement of the combined analysis is very small, compared with the exclusive 1 $b$-jet selection. In contrast with the results at $\mathcal{L}=139$ fb$^{-1}$, displayed in Fig.~\ref{2d-4q}, we see that one could increase the sensitivity on the Wilson coefficients of the four-quark operators by a factor $\sim 2.5$, reaching at 5$\sigma$ a scale of order $2.8$~TeV for unit coupling in $C_u$.

\begin{figure}[b!]
\begin{center}
\includegraphics[width=0.45\textwidth]{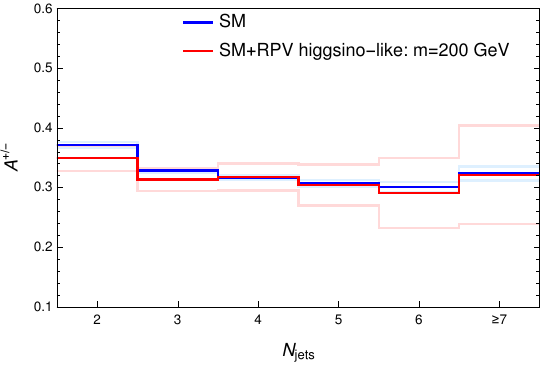}
\includegraphics[width=0.45\textwidth]{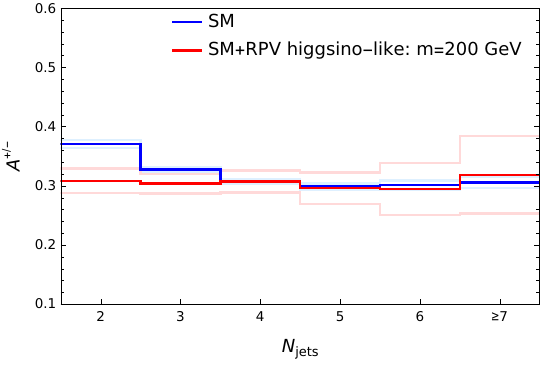}
\vspace{1mm}
\includegraphics[width=0.45\textwidth]{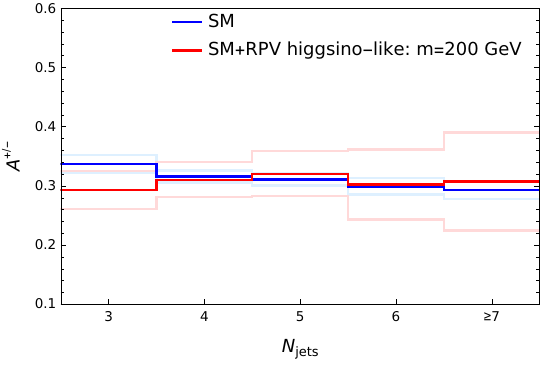}
\includegraphics[width=0.45\textwidth]{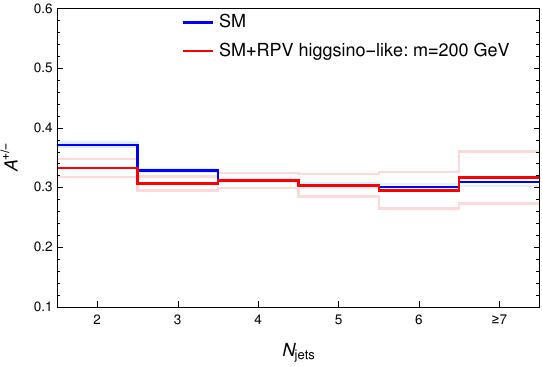}
\caption{Asymmetry vs $N_\text{jets}$, in the RPV-SUSY 200-GeV higgsino-like scenario at the HL-LHC. On the top left we show the exclusive 1 $b$-jet selection, on the top right the exclusive 2 $b$-jet selection, on the bottom left the inclusive 3 $b$-jet one and on the bottom right the inclusive 1 $b$-jet.}
\label{HLLHC-RPVNj}
\end{center}
\end{figure}

High luminosity LHC prospects for the RPV-SUSY 200-GeV higgsino-like scenario are shown in Fig.~\ref{HLLHC-RPVNj}. On the top left we show the exclusive 1 $b$-jet selection, on the top right the exclusive 2 $b$-jet selection, on the bottom left the inclusive 3 $b$-jet one and on the bottom right the 1 $b$-jet inclusive. There is an improvement with respect to the previous results at $\mathcal{L}=139$ fb$^{-1}$, although not enough to reach 5$\sigma$ in the higgsino-like case (but enough in the combined wino-like case). In fact, in the 200-GeV higgsino like case, that it is the most conservative case with the closest value to 1.5 $\times$ $\sigma_{t \bar t W}$ (it is indeed 1.54 $\times$ $\sigma_{t \bar t W}$), the significance increases by 2 to 3 orders of magnitude with respect to the inclusive selection with 1 $b$-jet at 139 fb$^{-1}$(see previous section). For the HL-LHC we obtain a significance for the inclusive 1 $b$-jet selection of 2.85, while for the alternative selections, we obtain 0.91 for the exclusive 1 $b$-jet selection, 2.99 for the inclusive 2 $b$-jet selection and 0.85 for the inclusive 3 $b$-jet selection, while the combined analysis provides a significance of 3.52. It should be noted that a more detailed study at 14 TeV would be necessary to probe if it is really possible to reach 5$\sigma$. It is also important to emphasize that although the combined analysis provides better results (as expected), the exclusive selection with 2 $b$-jets already improves by itself to the inclusive 1 $b$-jet (as opposed to the g2HDM case that there is no selection that by itself was better than the inclusive 1 $b$-jet selection). In the 400-GeV wino-like scenario, not shown here for the sake of brevity, for the inclusive 1 $b$-jet selection we obtain a significance of 4.43, although it should be important to clarify that is a scenario with 1.63 $\times$ $\sigma_{t \bar t W}$. For this case, for the alternative selections, we obtain significances of 0.40 for the exclusive 1 $b$-jet selection, 2.29 for the inclusive 2 $b$-jet selection and 4.37 for the inclusive 3 $b$-jet selection, while the combined analysis provides 5.15. Again the results of the combined analysis are better and the inclusive 3 $b$-jet selection alone, while not outperforming the inclusive 1 $b$-jet selection, is actually very good. These RPV-SUSY prospects suggest that it might be worth having a broader analysis of this model for HL-LHC, logically considering mass values that are allowed by the limits imposed by the most current searches as well as alternative scenarios with different coupling structure, including also alternative observables to the considered charge asymmetries.

\section{Conclusions}
\label{sec-conclusions}

In this article we have considered one of the most promising final-state signatures at the LHC: two same-sign leptons (electrons and/or muons) plus jets, with at least one tagged $b$-jet, that in the SM is dominated by $t\bar t W$, and receives contributions in many BSM theories. We have studied the charge asymmetry, defined as the difference between the yields of final states with two positive and two negative leptons, at the reconstructed level. We have analyzed the inclusive asymmetry, as well as differential distributions in terms of the following variables: the pseudorapidity and azimuthal angle difference between leptons, the lepton pair invariant mass, the scalar $p_T$, the sum of $H_T$ of the jets and of the leptons, the effective mass, the number of $b$-tagged jets and the number of jets. As benchmark BSM signals we have considered a general 2HDM, either with one of the spin zero states much lighter than the other, or with both of them at the same scale, exploring different masses and couplings; an effective theory with flavor violating four-quark operators, selecting for simplicity two operators with up and top quarks; and an $R$-parity violating supersymmetric model, exploring different masses of the unstable lightest supersymmetric particle, considered as higgsino- or wino-like.

For current LHC measurements at 13 TeV with a total integrated luminosity of ${\mathcal L}=139$~fb$^{-1}$, selecting the inclusive 1 tagged $b$-jet, we have determined the sensitivity to the detection of new physics in the BSM theories mentioned above by computing the $\chi^2$-distribution of the charge asymmetry as function of the number of jets. In the general 2HDM with the new scalars coupled to up- and top-quarks, considering different masses, we have studied the sensitivity as function of the generalized Yukawa couplings, determining that the size of the flavor violating coupling that could be tested to be around $\rho_{tu}\sim 0.1-0.8$ for $m_H\sim 200-1000$~GeV. In an effective theory with flavor violating four-quark operators, we have shown that a scale of order 2~TeV could be tested with the distributions of the charge asymmetry. On the other hand, the charge asymmetry does not seem to be a good observable to distinguish RPV supersymmetry from the SM, since by restricting the RPV-SUSY cross section to be smaller than 50\% of the $t\bar tW$ one, the distributions are within the error bands of the SM charge asymmetry and imply that the tests are not sensitive to the new physics. 

Finally we made a rough projection into the high luminosity regime of the LHC, by considering a total integrated luminosity of ${\mathcal L}=3000$~fb$^{-1}$ with a center-of-mass energy of $\sqrt{s}=13$~TeV, amounting to a reduction in the statistical uncertainty. Although the nominal expected energy for the high luminosity LHC is 14 TeV, this analysis provides conservative estimates of the high luminosity reach. We have considered different cuts in the $b$-jet selections: exclusive 1 and 2 $b$-jet selections, an inclusive 3 $b$-jet selection. For the general 2HDM we found that all selections are significantly above the SM, and that the combined analysis sensibly increases the sensitivity compared with the original selection at ${\mathcal L}=139$~fb$^{-1}$. For the effective theory with four-quark operators an exclusive 1 $b$-jet selection shows the largest sensitivity, while the improvement from other selections is negligible compared with this one. For RPV SUSY, in some cases it is possible to reach a significance of 5$\sigma$ in the analysis that combines the different selections, well above the case of ${\mathcal L}=139$~fb$^{-1}$. Our estimates point to interesting opportunities for the high luminosity and high energy LHC, though a more realistic analysis should be done at 13.6 TeV and 14 TeV, once we know the final energy at which the HL-LHC will operate. 

In summary, focusing on the 2SS$l$ plus jets final states and making use of some simple new physics scenarios, we have shown that the charge asymmetry is a powerful observable that, under current experimental and theory determinations of the SM dominating background of $t\bar{t}W$, can be used by differential measurements to probe new physics scenarios that potentially could contaminate the $t\bar{t}W$ signal.

\section*{Acknowledgments}
RMSS thanks Martin de los Rios for useful discussions. EA and RMSS acknowledge financial support from the ``Atracci\'on de Talento'' program (Modalidad 1) of the Comunidad de Madrid (Spain) under the grant number 2019-T1/TIC-14019, the Spanish Research Agency (Agencia Estatal de Investigaci\'on) through the Grant IFT Centro de Excelencia Severo Ochoa No CEX2020-001007-S and the grant PID2021-124704NB-I00 funded by MCIN/AEI/10.13039/501100011033. LD acknowledges support from PIP-11220200101426 and PICT-2017-2751. AJ was supported in part by the grants RTI2018-096930-B-I00 and PID2021-125273NB-I00. AM acknowledges support from CONICET and ANPCyT under project PICT 2018-03682.

\bibliography{multileptons}{}

\providecommand{\href}[2]{#2}\begingroup\raggedright\begin{thebibliography}{10}

\bibitem{ATLAS:2019nvo}
{ATLAS Collaboration}, \emph{{Analysis of $t\bar{t}H$ and $t\bar{t}W$
  production in multilepton final states with the ATLAS detector}},
  {\emph{\href{https://cds.cern.ch/record/2693930}{ATLAS-CONF-2019-045}} (2019)
  }.

\bibitem{CMS:2019rvj}
{CMS Collaboration}, \emph{{Search for production of four top quarks in final
  states with same-sign or multiple leptons in proton-proton collisions at
  $\sqrt{s}=$ 13 TeV}},
  \href{https://doi.org/10.1140/epjc/s10052-019-7593-7}{\emph{Eur. Phys. J. C}
  {\bfseries 80} (2020) 75}
  [\href{https://arxiv.org/abs/arXiv:1908.06463}{{\ttfamily
  arXiv:1908.06463}}].

\bibitem{CMS:2020mpn}
{CMS Collaboration}, \emph{{Measurement of the Higgs boson production rate in
  association with top quarks in final states with electrons, muons, and
  hadronically decaying tau leptons at $\sqrt{s} =$ 13 TeV}},
  \href{https://doi.org/10.1140/epjc/s10052-021-09014-x}{\emph{Eur. Phys. J. C}
  {\bfseries 81} (2021) 378}
  [\href{https://arxiv.org/abs/arXiv:2011.03652}{{\ttfamily
  arXiv:2011.03652}}].

\bibitem{ATLAS:2020hpj}
{ATLAS Collaboration}, \emph{{Evidence for $t\bar{t}t\bar{t}$ production in the
  multilepton final state in proton\textendash{}proton collisions at
  $\sqrt{s}=13$ $\text {TeV}$ with the ATLAS detector}},
  \href{https://doi.org/10.1140/epjc/s10052-020-08509-3}{\emph{Eur. Phys. J. C}
  {\bfseries 80} (2020) 1085}
  [\href{https://arxiv.org/abs/arXiv:2007.14858}{{\ttfamily
  arXiv:2007.14858}}].

\bibitem{ATLAS:2021kqb}
{ATLAS Collaboration}, \emph{{Measurement of the t$ \overline{t} $t$
  \overline{t} $ production cross section in $pp$ collisions at $ \sqrt{s} $ =
  13 TeV with the ATLAS detector}},
  \href{https://doi.org/10.1007/JHEP11(2021)118}{\emph{JHEP} {\bfseries 11}
  (2021) 118} [\href{https://arxiv.org/abs/arXiv:2106.11683}{{\ttfamily
  arXiv:2106.11683}}].

\bibitem{CMS:2023zdh}
{CMS Collaboration}, \emph{{Evidence for four-top quark production in
  proton-proton collisions at $\sqrt{s}$ = 13 TeV}},
  \href{https://doi.org/10.1016/j.physletb.2023.138076}{\emph{Phys. Lett. B}
  {\bfseries 844} (2023) 138076}
  [\href{https://arxiv.org/abs/arXiv:2303.03864}{{\ttfamily
  arXiv:2303.03864}}].

\bibitem{ATLAS:2023ajo}
{ATLAS Collaboration}, \emph{{Observation of four-top-quark production in the
  multilepton final state with the ATLAS detector}},
  \href{https://doi.org/10.1140/epjc/s10052-023-11573-0}{\emph{Eur. Phys. J. C}
  {\bfseries 83} (2023) 496}
  [\href{https://arxiv.org/abs/arXiv:2303.15061}{{\ttfamily
  arXiv:2303.15061}}].

\bibitem{CMS:2023ftu}
{CMS Collaboration}, \emph{{Observation of four top quark production in
  proton-proton collisions at $\sqrt{s}$ = 13 TeV}},
  \href{https://arxiv.org/abs/arXiv:2305.13439}{{\ttfamily arXiv:2305.13439}}.

\bibitem{ATLAS:2022rws}
{ATLAS Collaboration}, \emph{{Search for $ t\overline{t}H/A\to
  t\overline{t}t\overline{t} $ production in the multilepton final state in
  proton\textendash{}proton collisions at $ \sqrt{s} $ = 13 TeV with the ATLAS
  detector}}, \href{https://doi.org/10.1007/JHEP07(2023)203}{\emph{JHEP}
  {\bfseries 07} (2023) 203}
  [\href{https://arxiv.org/abs/arXiv:2211.01136}{{\ttfamily
  arXiv:2211.01136}}].

\bibitem{ATLAS:2023tlp}
{ATLAS Collaboration}, \emph{{Search for heavy Higgs bosons with
  flavour-violating couplings in multi-lepton plus $b$-jets final states in
  $pp$ collisions at 13 TeV with the ATLAS detector}},
  \href{https://arxiv.org/abs/arXiv:2307.14759}{{\ttfamily arXiv:2307.14759}}.

\bibitem{CMS:2020cpy}
{CMS Collaboration}, \emph{{Search for physics beyond the standard model in
  events with jets and two same-sign or at least three charged leptons in
  proton-proton collisions at $\sqrt{s}=$ 13 TeV}},
  \href{https://doi.org/10.1140/epjc/s10052-020-8168-3}{\emph{Eur. Phys. J. C}
  {\bfseries 80} (2020) 752}
  [\href{https://arxiv.org/abs/arXiv:2001.10086}{{\ttfamily
  arXiv:2001.10086}}].

\bibitem{ATLAS:2023lfr}
{ATLAS Collaboration}, \emph{{Search for direct production of winos and
  higgsinos in events with two same-charge leptons or three leptons in $pp$
  collision data at $\sqrt{s}=13$ TeV with the ATLAS detector}},
  \href{https://arxiv.org/abs/arXiv:2305.09322}{{\ttfamily arXiv:2305.09322}}.

\bibitem{CMS:2023xyc}
{CMS Collaboration}, \emph{{Search for physics beyond the standard model in top
  quark production with additional leptons in the context of effective field
  theory}},  \href{https://arxiv.org/abs/arXiv:2307.15761}{{\ttfamily
  arXiv:2307.15761}}.

\bibitem{LHCHiggsCrossSectionWorkingGroup:2016ypw}
{\scshape LHC Higgs Cross Section Working Group} collaboration, \emph{{Handbook
  of LHC Higgs Cross Sections: 4. Deciphering the Nature of the Higgs Sector}},
   \href{https://arxiv.org/abs/arXiv:1610.07922}{{\ttfamily arXiv:1610.07922}}.

\bibitem{Maltoni:2015ena}
F.~Maltoni, D.~Pagani and I.~Tsinikos, \emph{{Associated production of a
  top-quark pair with vector bosons at NLO in QCD: impact on $
  \mathrm{t}\overline{\mathrm{t}}\mathrm{H} $ searches at the LHC}},
  \href{https://doi.org/10.1007/JHEP02(2016)113}{\emph{JHEP} {\bfseries 02}
  (2016) 113} [\href{https://arxiv.org/abs/arXiv:1507.05640}{{\ttfamily
  arXiv:1507.05640}}].

\bibitem{Dror:2015nkp}
J.~A. Dror, M.~Farina, E.~Salvioni and J.~Serra, \emph{{Strong tW Scattering at
  the LHC}}, \href{https://doi.org/10.1007/JHEP01(2016)071}{\emph{JHEP}
  {\bfseries 01} (2016) 071}
  [\href{https://arxiv.org/abs/arXiv:1511.03674}{{\ttfamily
  arXiv:1511.03674}}].

\bibitem{Frederix:2017wme}
R.~Frederix, D.~Pagani and M.~Zaro, \emph{{Large NLO corrections in
  $t\bar{t}W^{\pm}$ and $t\bar{t}t\bar{t}$ hadroproduction from supposedly
  subleading EW contributions}},
  \href{https://doi.org/10.1007/JHEP02(2018)031}{\emph{JHEP} {\bfseries 02}
  (2018) 031} [\href{https://arxiv.org/abs/arXiv:1711.02116}{{\ttfamily
  arXiv:1711.02116}}].

\bibitem{Broggio:2019ewu}
A.~Broggio, A.~Ferroglia, R.~Frederix, D.~Pagani, B.~D. Pecjak and I.~Tsinikos,
  \emph{{Top-quark pair hadroproduction in association with a heavy boson at
  NLO+NNLL including EW corrections}},
  \href{https://doi.org/10.1007/JHEP08(2019)039}{\emph{JHEP} {\bfseries 08}
  (2019) 039} [\href{https://arxiv.org/abs/arXiv:1907.04343}{{\ttfamily
  arXiv:1907.04343}}].

\bibitem{Kulesza:2020nfh}
A.~Kulesza, L.~Motyka, D.~Schwartl\"ander, T.~Stebel and V.~Theeuwes,
  \emph{{Associated top quark pair production with a heavy boson: differential
  cross sections at NLO+NNLL accuracy}},
  \href{https://doi.org/10.1140/epjc/s10052-020-7987-6}{\emph{Eur. Phys. J. C}
  {\bfseries 80} (2020) 428}
  [\href{https://arxiv.org/abs/arXiv:2001.03031}{{\ttfamily
  arXiv:2001.03031}}].

\bibitem{Frederix:2020jzp}
R.~Frederix and I.~Tsinikos, \emph{{Subleading EW corrections and
  spin-correlation effects in $t\bar{t}W$ multi-lepton signatures}},
  \href{https://doi.org/10.1140/epjc/s10052-020-8388-6}{\emph{Eur. Phys. J. C}
  {\bfseries 80} (2020) 803}
  [\href{https://arxiv.org/abs/arXiv:2004.09552}{{\ttfamily
  arXiv:2004.09552}}].

\bibitem{Bevilacqua:2020pzy}
G.~Bevilacqua, H.-Y. Bi, H.~B. Hartanto, M.~Kraus and M.~Worek, \emph{{The
  simplest of them all: $t\bar{t} W^\pm$ at NLO accuracy in QCD}},
  \href{https://doi.org/10.1007/JHEP08(2020)043}{\emph{JHEP} {\bfseries 08}
  (2020) 043} [\href{https://arxiv.org/abs/arXiv:2005.09427}{{\ttfamily
  arXiv:2005.09427}}].

\bibitem{Denner:2020hgg}
A.~Denner and G.~Pelliccioli, \emph{{NLO QCD corrections to off-shell
  $\text{t}\bar{\text{t}}\text{W}^+$ production at the LHC}},
  \href{https://doi.org/10.1007/JHEP11(2020)069}{\emph{JHEP} {\bfseries 11}
  (2020) 069} [\href{https://arxiv.org/abs/arXiv:2007.12089}{{\ttfamily
  arXiv:2007.12089}}].

\bibitem{Bevilacqua:2020srb}
G.~Bevilacqua, H.-Y. Bi, H.~B. Hartanto, M.~Kraus, J.~Nasufi and M.~Worek,
  \emph{{NLO QCD corrections to off-shell ${t{\bar{t}}W^\pm }$ production at
  the LHC: correlations and asymmetries}},
  \href{https://doi.org/10.1140/epjc/s10052-021-09478-x}{\emph{Eur. Phys. J. C}
  {\bfseries 81} (2021) 675}
  [\href{https://arxiv.org/abs/arXiv:2012.01363}{{\ttfamily
  arXiv:2012.01363}}].

\bibitem{FebresCordero:2021kcc}
F.~Febres~Cordero, M.~Kraus and L.~Reina, \emph{{Top-quark pair production in
  association with a $W^\pm$ gauge boson in the POWHEG-BOX}},
  \href{https://doi.org/10.1103/PhysRevD.103.094014}{\emph{Phys. Rev. D}
  {\bfseries 103} (2021) 094014}
  [\href{https://arxiv.org/abs/arXiv:2101.11808}{{\ttfamily
  arXiv:2101.11808}}].

\bibitem{Frederix:2021agh}
R.~Frederix and I.~Tsinikos, \emph{{On improving NLO merging for $
  \mathrm{t}\overline{\mathrm{t}}\mathrm{W} $ production}},
  \href{https://doi.org/10.1007/JHEP11(2021)029}{\emph{JHEP} {\bfseries 11}
  (2021) 029} [\href{https://arxiv.org/abs/arXiv:2108.07826}{{\ttfamily
  arXiv:2108.07826}}].

\bibitem{Bevilacqua:2021tzp}
G.~Bevilacqua, H.~Y. Bi, F.~Febres~Cordero, H.~B. Hartanto, M.~Kraus, J.~Nasufi
  et~al., \emph{{Modeling uncertainties of $t\bar{t}W^\pm$ multilepton
  signatures}}, \href{https://doi.org/10.1103/PhysRevD.105.014018}{\emph{Phys.
  Rev. D} {\bfseries 105} (2022) 014018}
  [\href{https://arxiv.org/abs/arXiv:2109.15181}{{\ttfamily
  arXiv:2109.15181}}].

\bibitem{Buonocore:2023ljm}
L.~Buonocore, S.~Devoto, M.~Grazzini, S.~Kallweit, J.~Mazzitelli, L.~Rottoli
  et~al., \emph{{Associated production of a W boson with a top-antitop quark
  pair: next-to-next-to-leading order QCD predictions for the LHC}},
  \href{https://arxiv.org/abs/arXiv:2306.16311}{{\ttfamily arXiv:2306.16311}}.

\bibitem{ATLAS:2023gon}
{ATLAS Collaboration}, \emph{{Measurement of the total and differential
  cross-sections of $t\bar{t}W$ production in $pp$ collisions at 13 TeV with
  the ATLAS detector}},
  {\emph{\href{https://cds.cern.ch/record/2855337}{ATLAS-CONF-2023-019}} (2023)
  }.

\bibitem{CMS:2022tkv}
{\scshape CMS} collaboration, \emph{{Measurement of the cross section of top
  quark-antiquark pair production in association with a W boson in
  proton-proton collisions at $ \sqrt{s} $ = 13 TeV}},
  \href{https://doi.org/10.1007/JHEP07(2023)219}{\emph{JHEP} {\bfseries 07}
  (2023) 219} [\href{https://arxiv.org/abs/arXiv:2208.06485}{{\ttfamily
  arXiv:2208.06485}}].

\bibitem{Aaboud:2019njj}
{ATLAS Collaboration}, \emph{{Measurement of the $t\bar{t}Z$ and $t\bar{t}W$
  cross sections in proton-proton collisions at $\sqrt{s}=13$ TeV with the
  ATLAS detector}},
  \href{https://doi.org/10.1103/PhysRevD.99.072009}{\emph{Phys. Rev. D}
  {\bfseries 99} (2019) 072009}
  [\href{https://arxiv.org/abs/arXiv:1901.03584}{{\ttfamily
  arXiv:1901.03584}}].

\bibitem{Sirunyan:2017uzs}
{CMS Collaboration}, \emph{{Measurement of the cross section for top quark pair
  production in association with a W or Z boson in proton-proton collisions at
  $\sqrt{s} =$ 13 TeV}},
  \href{https://doi.org/10.1007/JHEP08(2018)011}{\emph{JHEP} {\bfseries 08}
  (2018) 011} [\href{https://arxiv.org/abs/arXiv:1711.02547}{{\ttfamily
  arXiv:1711.02547}}].

\bibitem{Frixione:2015zaa}
S.~Frixione, V.~Hirschi, D.~Pagani, H.~S. Shao and M.~Zaro, \emph{{Electroweak
  and QCD corrections to top-pair hadroproduction in association with heavy
  bosons}}, \href{https://doi.org/10.1007/JHEP06(2015)184}{\emph{JHEP}
  {\bfseries 06} (2015) 184}
  [\href{https://arxiv.org/abs/arXiv:1504.03446}{{\ttfamily
  arXiv:1504.03446}}].

\bibitem{Alwall:2014hca}
J.~Alwall, R.~Frederix, S.~Frixione, V.~Hirschi, F.~Maltoni, O.~Mattelaer
  et~al., \emph{{The automated computation of tree-level and next-to-leading
  order differential cross sections, and their matching to parton shower
  simulations}}, \href{https://doi.org/10.1007/JHEP07(2014)079}{\emph{JHEP}
  {\bfseries 07} (2014) 079}
  [\href{https://arxiv.org/abs/arXiv:1405.0301}{{\ttfamily arXiv:1405.0301}}].

\bibitem{Ball:2012cx}
R.~D. Ball et~al., \emph{{Parton distributions with LHC data}},
  \href{https://doi.org/10.1016/j.nuclphysb.2012.10.003}{\emph{Nucl. Phys. B}
  {\bfseries 867} (2013) 244}
  [\href{https://arxiv.org/abs/arXiv:1207.1303}{{\ttfamily arXiv:1207.1303}}].

\bibitem{Maltoni:2014zpa}
F.~Maltoni, M.~L. Mangano, I.~Tsinikos and M.~Zaro, \emph{{Top-quark charge
  asymmetry and polarization in $t\overline{t}W^\pm$ production at the LHC}},
  \href{https://doi.org/10.1016/j.physletb.2014.07.033}{\emph{Phys. Lett. B}
  {\bfseries 736} (2014) 252}
  [\href{https://arxiv.org/abs/1406.3262}{{\ttfamily 1406.3262}}].

\bibitem{Artoisenet:2012st}
P.~Artoisenet, R.~Frederix, O.~Mattelaer and R.~Rietkerk, \emph{{Automatic
  spin-entangled decays of heavy resonances in Monte Carlo simulations}},
  \href{https://doi.org/10.1007/JHEP03(2013)015}{\emph{JHEP} {\bfseries 03}
  (2013) 015} [\href{https://arxiv.org/abs/1212.3460}{{\ttfamily 1212.3460}}].

\bibitem{Frixione:2007zp}
S.~Frixione, E.~Laenen, P.~Motylinski and B.~R. Webber, \emph{{Angular
  correlations of lepton pairs from vector boson and top quark decays in Monte
  Carlo simulations}},
  \href{https://doi.org/10.1088/1126-6708/2007/04/081}{\emph{JHEP} {\bfseries
  04} (2007) 081} [\href{https://arxiv.org/abs/hep-ph/0702198}{{\ttfamily
  hep-ph/0702198}}].

\bibitem{Sjostrand:2014zea}
T.~Sj\"ostrand, S.~Ask, J.~R. Christiansen, R.~Corke, N.~Desai, P.~Ilten
  et~al., \emph{{An introduction to PYTHIA 8.2}},
  \href{https://doi.org/10.1016/j.cpc.2015.01.024}{\emph{Comput. Phys. Commun.}
  {\bfseries 191} (2015) 159}
  [\href{https://arxiv.org/abs/arXiv:1410.3012}{{\ttfamily arXiv:1410.3012}}].

\bibitem{Sjostrand:2007gs}
T.~Sjostrand, S.~Mrenna and P.~Z. Skands, \emph{{A Brief Introduction to PYTHIA
  8.1}}, \href{https://doi.org/10.1016/j.cpc.2008.01.036}{\emph{Comput. Phys.
  Commun.} {\bfseries 178} (2008) 852}
  [\href{https://arxiv.org/abs/arXiv:0710.3820}{{\ttfamily arXiv:0710.3820}}].

\bibitem{deFavereau:2013fsa}
J.~de~Favereau, C.~Delaere, P.~Demin, A.~Giammanco, V.~Lema{\^\i}tre,
  A.~Mertens et~al., \emph{{DELPHES 3, A modular framework for fast simulation
  of a generic collider experiment}},
  \href{https://doi.org/10.1007/JHEP02(2014)057}{\emph{JHEP} {\bfseries 02}
  (2014) 057} [\href{https://arxiv.org/abs/arXiv:1307.6346}{{\ttfamily
  arXiv:1307.6346}}].

\bibitem{Skands:2014pea}
P.~Skands, S.~Carrazza and J.~Rojo, \emph{{Tuning PYTHIA 8.1: the Monash 2013
  Tune}}, \href{https://doi.org/10.1140/epjc/s10052-014-3024-y}{\emph{Eur.
  Phys. J. C} {\bfseries 74} (2014) 3024}
  [\href{https://arxiv.org/abs/arXiv:1404.5630}{{\ttfamily arXiv:1404.5630}}].

\bibitem{Babu:2022ycv}
K.~S. Babu, R.~K. Barman, D.~Gon\c{c}alves and A.~Ismail, \emph{{Probing Lepton
  Number Violation and Majorana Nature of Neutrinos at the LHC}},
  \href{https://arxiv.org/abs/arXiv:2212.08025}{{\ttfamily arXiv:2212.08025}}.

\bibitem{Isidori:2010kg}
G.~Isidori, Y.~Nir and G.~Perez, \emph{{Flavor Physics Constraints for Physics
  Beyond the Standard Model}},
  \href{https://doi.org/10.1146/annurev.nucl.012809.104534}{\emph{Ann. Rev.
  Nucl. Part. Sci.} {\bfseries 60} (2010) 355}
  [\href{https://arxiv.org/abs/arXiv:1002.0900}{{\ttfamily arXiv:1002.0900}}].

\bibitem{ParticleDataGroup:2022pth}
{\scshape Particle Data Group} collaboration, \emph{{Review of Particle
  Physics}}, \href{https://doi.org/10.1093/ptep/ptac097}{\emph{PTEP} {\bfseries
  2022} (2022) 083C01}.

\bibitem{Lee:1973iz}
T.~D. Lee, \emph{{A Theory of Spontaneous T Violation}},
  \href{https://doi.org/10.1103/PhysRevD.8.1226}{\emph{Phys. Rev. D} {\bfseries
  8} (1973) 1226}.

\bibitem{Branco:2011iw}
G.~C. Branco, P.~M. Ferreira, L.~Lavoura, M.~N. Rebelo, M.~Sher and J.~P.
  Silva, \emph{{Theory and phenomenology of two-Higgs-doublet models}},
  \href{https://doi.org/10.1016/j.physrep.2012.02.002}{\emph{Phys. Rept.}
  {\bfseries 516} (2012) 1} [\href{https://arxiv.org/abs/1106.0034}{{\ttfamily
  1106.0034}}].

\bibitem{Trodden:1998qg}
M.~Trodden, \emph{{Electroweak baryogenesis: A Brief review}},  in \emph{{33rd
  Rencontres de Moriond: Electroweak Interactions and Unified Theories}},
  pp.~471--480, 1998, \href{https://arxiv.org/abs/hep-ph/9805252}{{\ttfamily
  hep-ph/9805252}}.

\bibitem{Haber:1984rc}
H.~E. Haber and G.~L. Kane, \emph{{The Search for Supersymmetry: Probing
  Physics Beyond the Standard Model}},
  \href{https://doi.org/10.1016/0370-1573(85)90051-1}{\emph{Phys. Rept.}
  {\bfseries 117} (1985) 75}.

\bibitem{Peccei:1977hh}
R.~D. Peccei and H.~R. Quinn, \emph{{CP Conservation in the Presence of
  Instantons}}, \href{https://doi.org/10.1103/PhysRevLett.38.1440}{\emph{Phys.
  Rev. Lett.} {\bfseries 38} (1977) 1440}.

\bibitem{Kim:1986ax}
J.~E. Kim, \emph{{Light Pseudoscalars, Particle Physics and Cosmology}},
  \href{https://doi.org/10.1016/0370-1573(87)90017-2}{\emph{Phys. Rept.}
  {\bfseries 150} (1987) 1}.

\bibitem{Mrazek:2011iu}
J.~Mrazek, A.~Pomarol, R.~Rattazzi, M.~Redi, J.~Serra and A.~Wulzer, \emph{{The
  Other Natural Two Higgs Doublet Model}},
  \href{https://doi.org/10.1016/j.nuclphysb.2011.07.008}{\emph{Nucl. Phys. B}
  {\bfseries 853} (2011) 1} [\href{https://arxiv.org/abs/1105.5403}{{\ttfamily
  1105.5403}}].

\bibitem{Davidson:2005cw}
S.~Davidson and H.~E. Haber, \emph{{Basis-independent methods for the
  two-Higgs-doublet model}},
  \href{https://doi.org/10.1103/PhysRevD.72.099902}{\emph{Phys. Rev. D}
  {\bfseries 72} (2005) 035004}
  [\href{https://arxiv.org/abs/arXiv:hep-ph/0504050}{{\ttfamily
  arXiv:hep-ph/0504050}}].

\bibitem{Altunkaynak:2015twa}
B.~Altunkaynak, W.-S. Hou, C.~Kao, M.~Kohda and B.~McCoy, \emph{{Flavor
  Changing Heavy Higgs Interactions at the LHC}},
  \href{https://doi.org/10.1016/j.physletb.2015.10.024}{\emph{Phys. Lett. B}
  {\bfseries 751} (2015) 135}
  [\href{https://arxiv.org/abs/arXiv:1506.00651}{{\ttfamily
  arXiv:1506.00651}}].

\bibitem{Hou:2020ciy}
W.-S. Hou, T.-H. Hsu and T.~Modak, \emph{{Constraining the $t \to u$ flavor
  changing neutral Higgs coupling at the LHC}},
  \href{https://doi.org/10.1103/PhysRevD.102.055006}{\emph{Phys. Rev. D}
  {\bfseries 102} (2020) 055006}
  [\href{https://arxiv.org/abs/arXiv:2008.02573}{{\ttfamily
  arXiv:2008.02573}}].

\bibitem{Hou:2020chc}
W.-S. Hou and T.~Modak, \emph{{Probing Top Changing Neutral Higgs Couplings at
  Colliders}}, \href{https://doi.org/10.1142/S0217732321300068}{\emph{Mod.
  Phys. Lett. A} {\bfseries 36} (2021) 2130006}
  [\href{https://arxiv.org/abs/arXiv:2012.05735}{{\ttfamily
  arXiv:2012.05735}}].

\bibitem{Degrande:2014vpa}
C.~Degrande, \emph{{Automatic evaluation of UV and R2 terms for beyond the
  Standard Model Lagrangians: a proof-of-principle}},
  \href{https://doi.org/10.1016/j.cpc.2015.08.015}{\emph{Comput. Phys. Commun.}
  {\bfseries 197} (2015) 239}
  [\href{https://arxiv.org/abs/arXiv:1406.3030}{{\ttfamily arXiv:1406.3030}}].

\bibitem{ATLAS:2021uiz}
{ATLAS Collaboration}, \emph{{Search for resonances decaying into photon pairs
  in 139 fb$^{-1}$ of $pp$ collisions at $\sqrt {s}$=13 TeV with the ATLAS
  detector}}, \href{https://doi.org/10.1016/j.physletb.2021.136651}{\emph{Phys.
  Lett. B} {\bfseries 822} (2021) 136651}
  [\href{https://arxiv.org/abs/arXiv:2102.13405}{{\ttfamily
  arXiv:2102.13405}}].

\bibitem{CMS:2019pzc}
{CMS Collaboration}, \emph{{Search for heavy Higgs bosons decaying to a top
  quark pair in proton-proton collisions at $\sqrt{s} =$ 13 TeV}},
  \href{https://doi.org/10.1007/JHEP04(2020)171}{\emph{JHEP} {\bfseries 04}
  (2020) 171} [\href{https://arxiv.org/abs/arXiv:1908.01115}{{\ttfamily
  arXiv:1908.01115}}].

\bibitem{Arganda:2021yms}
E.~Arganda, L.~Da~Rold, D.~A. D\'\i{}az and A.~D. Medina, \emph{{Interpretation
  of LHC excesses in ditop and ditau channels as a 400-GeV pseudoscalar
  resonance}}, \href{https://doi.org/10.1007/JHEP11(2021)119}{\emph{JHEP}
  {\bfseries 11} (2021) 119}
  [\href{https://arxiv.org/abs/arXiv:2108.03058}{{\ttfamily
  arXiv:2108.03058}}].

\bibitem{Buchmuller:1985jz}
W.~Buchmuller and D.~Wyler, \emph{{Effective Lagrangian Analysis of New
  Interactions and Flavor Conservation}},
  \href{https://doi.org/10.1016/0550-3213(86)90262-2}{\emph{Nucl. Phys. B}
  {\bfseries 268} (1986) 621}.

\bibitem{Barbier:2004ez}
R.~Barbier et~al., \emph{{R-parity violating supersymmetry}},
  \href{https://doi.org/10.1016/j.physrep.2005.08.006}{\emph{Phys. Rept.}
  {\bfseries 420} (2005) 1}
  [\href{https://arxiv.org/abs/arXiv:hep-ph/0406039}{{\ttfamily
  arXiv:hep-ph/0406039}}].

\bibitem{Dreiner:2023bvs}
H.~K. Dreiner, Y.~S. Koay, D.~K\"ohler, V.~M. Lozano, J.~Montejo~Berlingen,
  S.~Nangia et~al., \emph{{The ABC of RPV: classification of R-parity violating
  signatures at the LHC for small couplings}},
  \href{https://doi.org/10.1007/JHEP07(2023)215}{\emph{JHEP} {\bfseries 07}
  (2023) 215} [\href{https://arxiv.org/abs/2306.07317}{{\ttfamily
  2306.07317}}].

\bibitem{Deshpande:2016yrv}
N.~G. Deshpande and X.-G. He, \emph{{Consequences of R-parity violating
  interactions for anomalies in $\bar B\to D^{(*)} \tau \bar \nu$ and $b\to s
  \mu^+\mu^-$}},
  \href{https://doi.org/10.1140/epjc/s10052-017-4707-y}{\emph{Eur. Phys. J. C}
  {\bfseries 77} (2017) 134}
  [\href{https://arxiv.org/abs/arXiv:1608.04817}{{\ttfamily
  arXiv:1608.04817}}].

\bibitem{Das:2017kfo}
D.~Das, C.~Hati, G.~Kumar and N.~Mahajan, \emph{{Scrutinizing $R$-parity
  violating interactions in light of $R_{K^{(\ast)}}$ data}},
  \href{https://doi.org/10.1103/PhysRevD.96.095033}{\emph{Phys. Rev. D}
  {\bfseries 96} (2017) 095033}
  [\href{https://arxiv.org/abs/arXiv:1705.09188}{{\ttfamily
  arXiv:1705.09188}}].

\bibitem{Earl:2018snx}
K.~Earl and T.~Gr\'egoire, \emph{{Contributions to ${b \rightarrow s \ell
  \ell}$ Anomalies from ${R}$-Parity Violating Interactions}},
  \href{https://doi.org/10.1007/JHEP08(2018)201}{\emph{JHEP} {\bfseries 08}
  (2018) 201} [\href{https://arxiv.org/abs/arXiv:1806.01343}{{\ttfamily
  arXiv:1806.01343}}].

\bibitem{Hu:2019ahp}
Q.-Y. Hu and L.-L. Huang, \emph{{Explaining $b\to s \ell^+ \ell^-$ data by
  sneutrinos in the $R$ -parity violating MSSM}},
  \href{https://doi.org/10.1103/PhysRevD.101.035030}{\emph{Phys. Rev. D}
  {\bfseries 101} (2020) 035030}
  [\href{https://arxiv.org/abs/arXiv:1912.03676}{{\ttfamily
  arXiv:1912.03676}}].

\bibitem{Hu:2020yvs}
Q.-Y. Hu, Y.-D. Yang and M.-D. Zheng, \emph{{Revisiting the $B$-physics
  anomalies in $R$-parity violating MSSM}},
  \href{https://doi.org/10.1140/epjc/s10052-020-7940-8}{\emph{Eur. Phys. J. C}
  {\bfseries 80} (2020) 365}
  [\href{https://arxiv.org/abs/arXiv:2002.09875}{{\ttfamily
  arXiv:2002.09875}}].

\bibitem{Heinemeyer:2021opc}
S.~Heinemeyer, E.~Kpatcha, I.~n. Lara, D.~E. L\'opez-Fogliani, C.~Mu\~noz and
  N.~Nagata, \emph{{The new $(g-2)_\mu $ result and the $\mu \nu $SSM}},
  \href{https://doi.org/10.1140/epjc/s10052-021-09601-y}{\emph{Eur. Phys. J. C}
  {\bfseries 81} (2021) 802}
  [\href{https://arxiv.org/abs/arXiv:2104.03294}{{\ttfamily
  arXiv:2104.03294}}].

\bibitem{Zheng:2021wnu}
M.-D. Zheng and H.-H. Zhang, \emph{{Studying the $b\rightarrow s \ell^+\ell^-$
  anomalies and $(g-2)_{\mu}$ in $R$-parity violating MSSM framework with the
  inverse seesaw mechanism}},
  \href{https://doi.org/10.1103/PhysRevD.104.115023}{\emph{Phys. Rev. D}
  {\bfseries 104} (2021) 115023}
  [\href{https://arxiv.org/abs/arXiv:2105.06954}{{\ttfamily
  arXiv:2105.06954}}].

\bibitem{Choudhury:2023lbp}
A.~Choudhury, S.~Mitra, A.~Mondal and S.~Mondal, \emph{{Bilinear R-parity
  violating supersymmetry under the light of neutrino oscillation, higgs and
  flavor data}},  \href{https://arxiv.org/abs/arXiv:2305.15211}{{\ttfamily
  arXiv:2305.15211}}.

\end{thebibliography}\endgroup
\bibliographystyle{JHEP}

\end{document}